\documentclass[pre,reprint]{revtex4-1}

\usepackage[lofdepth,lotdepth,caption=false]{subfig}
\usepackage{amsmath}
\usepackage{amssymb}
\usepackage{mathtools}
\usepackage{graphicx}
\usepackage{dcolumn}
\usepackage{color}

\begin{document}
\title{Invariant embedding approach to secondary electron emission from metals}
\author{F. X. Bronold and H. Fehske}
\date{\today}

\address{Institut f{\"u}r Physik,
        Universit{\"a}t Greifswald, 17489 Greifswald, Germany }
\begin{abstract}
Based on an invariant embedding principle for the backscattering function we calculate 
the electron emission yield for metal surfaces at very low electron impact energies. 
Solving the embedding equation within a quasi-isotropic approximation and using the 
effective mass model for the solid, experimental data are fairly well reproduced
provided (i) incoherent scattering on ion cores is allowed to contribute to the 
scattering cascades inside the solid and (ii) the transmission through the surface potential 
takes into account Bragg gaps due to coherent scattering on crystal planes parallel to the 
surface as well as randomization of the electron's lateral momentum due to elastic scattering 
on surface defects. Our results suggest that in order to get secondary electrons out of metals,
the large energy loss due to inelastic electron-electron scattering has to be compensated for
by incoherent elastic electron-ion core scattering, irrespective of the crystallinity of the 
sample. 
\end{abstract}

\maketitle

\section{Introduction}

Secondary electron emission upon electron impact has been investigated ever since the
start of modern solid state physics~\cite{Dekker58}. It is at the core of many technological
applications, for good and for bad, ranging from surface analysis~\cite{SW13,Cazaux12} 
over materials processing~\cite{KBK14} and electron storage rings, where electron 
multipacting due to secondary electrons may limit the ring's performance~\cite{CCF04}, 
to various plasma devices~\cite{TKR20,VDT18,CU16,Tolias14a,Tolias14b,CIL07,RSK05,TLC04}. 
Hence, a great number of experimental and theoretical work has been done in this field 
(for references to recent work see, for instance, the surveys in~\cite{BPR20,WEA08}). Most
of it is concerned with electron emission due to primary electrons hitting the surfaces 
at high energies, above $100\,\mathrm{eV}$~\cite{WEA08}. In plasma devices, however, 
which are our main interest, the primaries triggering secondary electrons typically 
have energies below $50\,\mathrm{eV}$. For instance, in dc microdischarges~\cite{CIL07}, 
the electron distribution functions close to the electrodes have substantial weight below 
$50\,\mathrm{eV}$. Most of the electrons hitting the walls confining the plasma have 
thus energies in a range where collective phenomena of the solid, such as diffraction, 
the surface potential, and polarization effects, start to play a role. 

The purpose of this work is to present an approach for calculating the emission yield which
accounts for these phenomena and to apply it to metal surfaces at very low electron impact 
energies. Instead of using kinetic equations of the Boltzmann-type~\cite{Wolff54,PAG85} 
or Monte-Carlo simulations of the process~\cite{AAC19,CAM18,PIB17,RJT13}, we employ the 
invariant embedding principle. Originally developed by Ambartsumian~\cite{Ambartsumian42}
and Chandrasekhar~\cite{Chandrasekhar60} for the description of radiative transport in 
stellar atmospheres, Dashen~\cite{Dashen64} has shown quite some time ago that it can be 
also used to analyze electron backscattering from solid surfaces. Various groups 
pursued since then the principle~\cite{ARK20,FBH14,GP07,GT03,Vicanek99,AP94}. But its full
potential has not been tapped yet because the applications were either restricted to 
elastic scattering or to radiative transfer. 

In previous studies, we employed the principle to calculate the sticking~\cite{BF15} and 
reflection~\cite{BF17a} probabilities for an electron hitting a dielectric surface. To avoid
electron-electron scattering, leading to electron multiplication due to the two final states, 
we restricted however the considerations to impact energies below the band gap. By adopting 
the reasoning of Wolff~\cite{Wolff54}, later refined by Penn and coworkers~\cite{PAG85}, we 
now enable our approach to treat also electron-electron scattering.

Besides generalizing conceptually our previous approach, we also develop in this work a
numerical scheme, based on a quasi-isotropic approximation for the angular integrals, 
which solves the nonlinear embedding equation for the electron backscattering function 
(and not just its linearization~\cite{BF15,BF17a}), while keeping the scattering processes
inelastic as well as angle-dependent. Analyzing within this scheme experimental data 
for various metals, we also hope to clarify two issues, raised specifically
by Cimino and coworkers' experiments on clean and as-received noble metal 
surfaces~\cite{CCF04,CGL15,GAL17}, which led in the plasma physics community to a 
debated~\cite{ASK13,Andronov14} revival of interest in low-energy electron backscattering 
from surfaces: (i) The discrepancy between the recent~\cite{CCF04,CGL15,GAL17} and  
previous measurements by Bronshtein and Roshchin~\cite{BR58} and (ii) the increase of the 
emission yields to unity for vanishing impact energy found in the recent data. 

Using an effective mass model, augmented by electron scattering on the ion cores,
Bragg diffraction on the crystal planes parallel to the surface, and scattering 
on surface defects, we find rather good agreement with measured 
data~\cite{GAL17,MC76,YG72,YG70,BronFrai69,KHA63,BR58}. Except for the scattering
strength of the surface defects, the model contains no free parameters. From a broader
perspective, our results suggest (i) that electron-electron scattering alone yields 
essentially no secondary electrons, the emission yields being given in that case by 
the electron reflectivities of the metal surfaces, which, in the absence of Bragg gaps, 
are very small due to the image charge effect, (ii) the potentials of the ion cores 
act as incoherent scattering centers, irrespective of the crystallinity of the sample,
implying that Bauer's randium model~\cite{Bauer70}, often used at high impact energies, 
may be a better starting point for the theoretical analysis of low energy data as one 
would perhaps expect, and (iii) inclusion of surface imperfections in the surface transmission
function may be also required for a quantitative description of electron emission at 
very low impact energies.

The outline of the paper is as follows. In Section~\ref{Method} we present our approach.
It is divided into three subsections, introducing the expression for the emission yield,
the definition of the backscattering function together with its determination via the 
embedding equation, and the augmented effective mass model for the solid. Section~\ref{Results}
presents numerical results for Al, W, and the noble metals Cu, Ag, and Au. Concluding
remarks are given in Section~\ref{Conclusions} while technical details interrupting 
the flow of the presentation are given in three appendices. 

\section{Calculational approach}
\label{Method}

\subsection{Electron emission yield}
\label{MethodYield}

To calculate the electron emission yield due to a primary electron hitting the 
metal surface with an energy $E$ and an external direction cosine $\xi=\cos\beta$, 
we image the electron to first encounter the surface potential of depth 
$U=E_\mathrm{F}+\Phi$, where $E_\mathrm{F}$ and $\Phi$ are the Fermi energy and 
work function of the metal. For an idealized surface, with perfect homogeneity in 
the lateral directions, the azimuth angles can be integrated out. Only the polar 
angles thus enter the formalism, giving rise to the direction cosines. In case the 
electron successfully traverses the potential, it may initiate scattering cascades 
inside the metal leading to electrons which in turn may escape the solid if they are 
directed towards the interface and have enough kinetic energy in the perpendicular 
motion to traverse the surface potential from the inside out.
 
The emission yield $Y(E,\xi)$ may thus be expressed as (throughout we measure energy in 
Rydbergs, length in Bohr radii, and mass in bare electron masses)
\begin{align}
Y(E,\xi)=R(E,\xi) + \big(1-R(E,\xi)\big){\cal E}(E,\xi)~,
\label{EEY}
\end{align}
where $R(E,\xi)$ is the reflection probability due to the metal's surface potential and 
\begin{align}
{\cal E}(E,\xi)=\!\int_{\eta_{\rm min}(E)}^1 \!\!\!\!\!\!\!\!\!d\eta^\prime 
\int_{E_{\rm min}(\eta^\prime)}^E \!\!\!\!\!\!\!\!\!\!\!\!\!\!\!\!dE^\prime
\rho(E^\prime) {\cal B}(E\eta(\xi)|E^\prime\eta^\prime)D(E^\prime,\xi(\eta^\prime))
\label{EscapeFct}
\end{align}
is the escape function, describing the emission of secondary electrons due to the 
scattering cascades inside the metal. It contains the density of states $\rho(E)$ of the 
metal's conduction band, the transmission probability 
\begin{align}
D(E,\xi)=1-R(E,\xi)~,
\end{align}
and the function ${\cal B}(E\eta|E^\prime\eta^\prime)$, encoding backscattering cascades 
from an initial electron state $(E,\eta)$ to a final electron state $(E^\prime,\eta^\prime)$. 
The functions $\xi(\eta)$ and $\eta(\xi)$ connect internal and external direction cosines 
and are implicitly defined by the relation
\begin{align}
1-\eta^2=(1-\xi^2)\frac{E}{E+U}~,
\end{align} 
where we assumed (as for all formulae presented below) a quadratic dispersion for the 
conduction band of the metal, with an effective mass equal to the bare electron mass. 
The relation follows from the conservation of energy and lateral momentum at the interface. 
Constraining the integrations over $\eta^\prime$ and $E^\prime$ from below by 
$\eta_\mathrm{min}(E)=\sqrt{U/(E+U)}$ and $E_\mathrm{min}(\eta^\prime)=U(1/\eta^{\prime\,2}-1)$ 
ensures that only backscattered electrons with a perpendicular kinetic energy larger than the 
depth of the surface potential contribute to the emission yield. For the quadratic dispersion,
finally, $\rho(E)=\sqrt{(E+U)}/2$.

\begin{figure*}[t]
\includegraphics[width=0.90\linewidth]{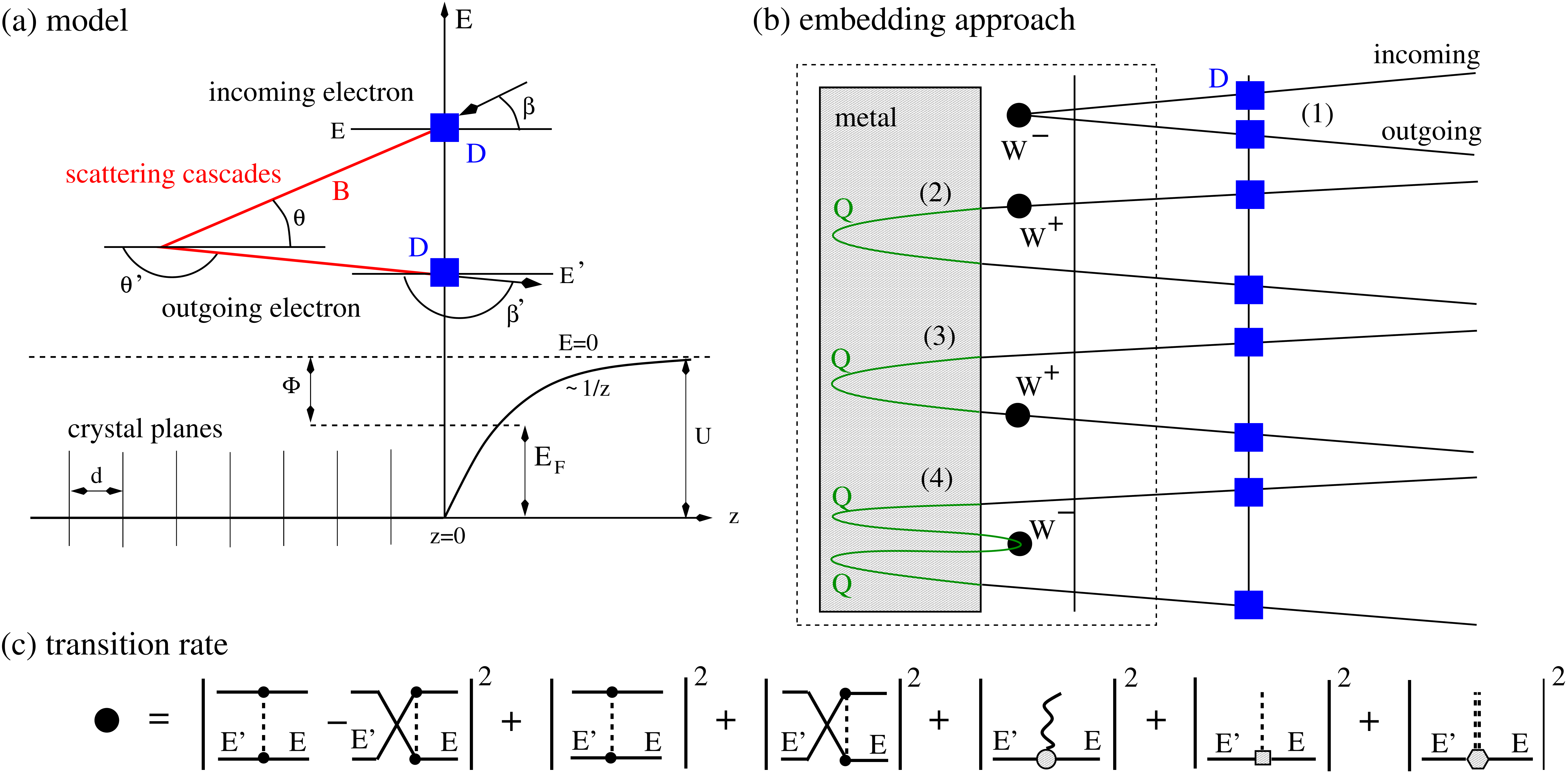}
\caption{(color online)
(a) Illustration of the main parts of the surface model. (b) Principle of
invariant embedding as used to determine the function ${\cal B}(E\eta|E^\prime\eta^\prime)$
into which the scattering cascades are encoded and (c) diagrammatic representation of
the processes included in the transition rate $W^\pm(E\eta|E^\prime\eta^\prime)$. The first
three terms are due to electron-electron scattering, while the following ones are due 
to electron-phonon, electron-impurity, and electron-ion core scattering. The
labelling of the diagrams is not complete. It is only meant to indicate where in the 
diagrams the initial and final states appear.  
}
\label{MethodCartoon}
\end{figure*}

The scattering geometry and the structure of Eqs.~\eqref{EEY} and~\eqref{EscapeFct} are 
visualized in Fig.~\ref{MethodCartoon}a. From it, the definition of the direction cosines 
can be also inferred, $\eta^\prime=|\cos\theta^\prime|$ with $\pi/2 \le \theta^\prime \le \pi$
and $\eta=\cos\theta$ with $0 \le \theta\le \pi/2$. Similar relations hold for the external 
direction cosines $\xi$ and $\xi^\prime$ and their corresponding angles. The surface potential
we use for the calculation of the transmission function $D(E,\xi)$ is also indicated. In 
addition to the image tail, we allow for Bragg scattering on crystal planes parallel to 
the interface. Before we describe how we calculate the functions entering our approach, we 
generalize it to an imperfect interface, where electrons may scatter elastically on surface 
defects while passing the interface. 

At an imperfect surface, lateral momentum is not conserved due to lack of in-plane homogeneity. 
To take this possibility into account we employ an approach which was originally developed 
by Smith and coworkers~\cite{SLN98} to analyze ballistic electron spectroscopy data for 
semiconductor-metal interfaces. Later, we used it to study the interaction of an electron
with imperfect dielectric surfaces~\cite{BF15,BF17a}. Adopting the 
notation to the case of metals, the emission yield for an imperfect metal surface becomes,
\begin{align}
\overline{Y}(E,\xi)=1-\overline{S}(E,\xi)~,
\label{Ybar}
\end{align}
where
\begin{align}
\overline{S}(E,\xi)&=\frac{D(E,\xi)}{1+C/\xi}\big[1-\overline{\cal E}(E,\xi)\big] \nonumber\\
         &+ \frac{C/\xi}{1+C/\xi}\int^1_0 \!\!\!d\xi^\prime D(E,\xi^\prime)
         \big[1-\overline{\cal E}(E,\xi^\prime)\big]
\end{align}
is the probability (strictly speaking, quasi-probability, see next subsection) for not emitting 
an electron. The function $\overline{\cal E}(E,\xi)$ entering 
this expression is given by~\eqref{EscapeFct} with $D(E,\xi)$ in the integrand replaced by 
\begin{align}
\overline{D}(E,\xi)=\frac{D(E,\xi)}{1+C/\xi}
        + \frac{C/\xi}{1+C/\xi}\int^1_0 d\xi^\prime D(E,\xi^\prime)~.
\label{Dbar}
\end{align}

The function $\overline{D}(E,\xi)$ describes the transmission through the surface potential 
in the presence of elastic scattering on defects. It causes relaxation of the lateral 
momentum. Diffuse transmission (second term in~\eqref{Dbar}) is thus possible at expense 
of ballistic transmission (first term in~\eqref{Dbar}). The scattering strength $C$ is 
proportional to the product of the absolute square of the matrix element and the defect 
concentration. Both are unknown. We thus take $C$ as an adjustable parameter.
The emission yield of the perfect surface, Eq.~\eqref{EEY}, is recovered from 
Eqs.~\eqref{Ybar}--\eqref{Dbar} in the weak scattering limit $C\ll 1$, while the yield of 
a rather dirty, irregular surface is given in the strong scattering limit $C\gg 1$.

\subsection{Backscattering function}
\label{MethodBackScattSect}

The scattering cascades inside the metal are encoded in the function ${\cal B}(E\eta|E^\prime\eta^\prime)$,
which in turn is related to the backscattering function $Q(E\eta|E^\prime\eta^\prime)$ which, following 
the work of Dashen~\cite{Dashen64}, is obtained from the invariant embedding principle depicted 
in Fig.~\ref{MethodCartoon}b. The principle states that adding an infinitesimally thin layer of the same 
material to a halfspace already filled by it does not change the backscattering. It leads to the 
embedding equation for $Q(E\eta|E^\prime\eta^\prime)$. Decoupling angle and energy variables by  
the quasi-isotropic approximation described in Appendix~\ref{AppendixA}, the embedding equation reads
\begin{widetext}
\begin{align}
Q(E|E^\prime;\eta\eta^\prime) &= K^-(E\eta|E^\prime\eta^\prime)
+ \int_{E^\prime}^E \!\!dE^{\prime\prime} K^+_1(E|E^{\prime\prime};E^\prime\eta\eta^\prime)
Q(E^{\prime\prime}|E^\prime;\eta\eta^\prime)
+\int_{E^\prime}^E \!\!dE^{\prime\prime} Q(E|E^{\prime\prime};\eta\eta^\prime)
K^+_2(E^{\prime\prime}|E^\prime;E\eta\eta^\prime)\nonumber\\
&+ \int_{E^\prime}^E \!\!dE^{\prime\prime} \!\!\int_{E^\prime}^{E^{\prime\prime}}\!\!dE^{\prime\prime\prime}
Q(E|E^{\prime\prime};\eta\eta^\prime)
B^-(E^{\prime\prime}|E^{\prime\prime\prime};E E^\prime \eta\eta^\prime)
Q(E^{\prime\prime\prime}|E^\prime;\eta\eta^\prime)~,
\label{EmbeddingEq}
\end{align}
\end{widetext}
where the notation $Q(E|E^\prime;\eta\eta^\prime)$ signals the different status of the energy and angle 
variables, with only the former affected by the integrations and the latter simply external 
parameters.

The validity of the decoupling hinges on the angular dependence of $K^-(E\eta|E^\prime\eta^\prime)$.
It should be nearly isotropic. In this sense, the decoupling is closely related to the transport 
approximation employed by Werner and coworkers in their analysis of elastic electron backscattering 
from surfaces~\cite{WTH94}. Using the surface model described in the next subsection, we show in 
Appendix~\ref{AppendixA} data for $K^-(E\eta|E^\prime\eta^\prime)$ which suggest that, for elastic
scattering the approximation is indeed well justified, while for inelastic electron-electron 
collisions it becomes somewhat questionable, especially for large direction cosines. However, the rather 
good agreement between calculated and measured emission yields supports--in retrospect--the assumption 
that, as a first step towards solving the full embedding equation, the angular dependence of 
$K^-(E\eta|E^\prime\eta^\prime)$ can be considered nearly isotropic, irrespective of the energy 
transfer. Details of the decoupling and the definition of the functions $K^-$, $K_1^+$, $K_2^+$, 
and $B^-$ are given in Appendix~\ref{AppendixA}. 

Separating inelastic electron-electron from elastic collisions by writing the functions 
entering Eq.~\eqref{EmbeddingEq} as a sum of two terms, 
\begin{align}
A(E|E^\prime;\eta\eta^\prime)=A_{\rm ee}(E|E^\prime;\eta\eta^\prime)+A_{\rm el}(E;\eta\eta^\prime)\delta(E-E^\prime)~,
\label{Splitting}
\end{align}
and expanding the inelastic part of the backscattering function in the number of backscattering events, 
\begin{align}
Q(E|E^\prime;\eta\eta^\prime) &=\sum_{l=0}^{n_\mathrm{max}} Q_{\rm ee}^{(2l+1)}(E|E^\prime;\eta\eta^\prime)
\nonumber\\
&+Q_{\rm el}(E;\eta\eta^\prime)\delta(E-E^\prime)~,
\label{Qexpansion}
\end{align}
enables us to solve the embedding equation by an iterative process which successively increases the 
number of inelastic backscattering events. The functions $Q_{\rm ee}^{(2l+1)}(E|E^\prime;\eta\eta^\prime)$ turn
out to satisfy a set of linear integral equations with kernels renormalized by $Q_{\rm el}(E;\eta\eta^\prime)$, 
which itself is given by the positive solution of a quadratic algebraic equation. Essential for the 
feasibility of the approach is the Volterra-type structure of the energy integrals in the embedding 
equation. It allows to sweep the $EE^\prime$-plane in such a manner that the $Q_{\rm ee}^{(k)}$ 
appearing in the kernels of the integral equation for $Q_{\rm ee}^{(n)}$ with $n>k$ are given from 
the previous steps. The renormalized kernels and further details about our strategy to 
solve~\eqref{EmbeddingEq} are given in Appendix~\ref{AppendixB}.

To obtain the function ${\cal B}(E\eta|E^\prime\eta^\prime)$ entering~\eqref{EscapeFct}, we have 
to keep in mind that not all backscattered electrons leave the solid. We also have to take 
into account that in an electron-electron scattering event, the initial electron leads to two final
electrons. In terms of the two contributions to the backscattering function, $Q_{\rm ee}$ and 
$Q_{\rm el}$, we therefore write 
\begin{align}
{\cal B}(E\eta|E^\prime\eta^\prime)=
\frac{2 Q_{\rm ee}(E|E^\prime;\eta\eta^\prime)+Q_{\rm el}(E^\prime;\eta\eta^\prime)\delta(E-E^\prime)}
{\int_{-\Phi}^E dE^{\prime\prime}\int_{\eta_c}^1 d\eta^{\prime\prime}\rho(E^{\prime\prime})
Q(E|E^{\prime\prime};\eta\eta^{\prime\prime})}~,
\label{Bfct}
\end{align}
where the factor two in front of $Q_{\rm ee}$ in the numerator accounts for the two final electrons  
in an electron-electron backscattering event~\cite{PAG85,Wolff54} and the normalization ensures that 
--at the end in Eq.~\eqref{EscapeFct}--only backscattered electrons contribute to the emission yield 
which make it also over the potential barrier and thus leave the metal. Hence, with the function 
${\cal B}(E\eta|E^\prime\eta^\prime)$ defined in~\eqref{Bfct}, the escape function~\eqref{EscapeFct} 
formally resembles a conditional (pseudo-)probability (without electron-electron collisions, that is, 
without electron multiplication, it would be a probability in the strict sense).

The treatment so far did not use any properties of the transition rates associated with the 
elementary scattering processes. Only the angular dependencies are postulated to behave in a 
manner to justify the decoupling of the energy and angle variables. To furnish the equations, 
a model for the solid is required. 

\subsection{Augmented effective mass model}
\label{MethodModel}

To obtain numerical results we need a model for
the surface potential and a model for the bulk scattering processes. The former enters
the calculation of the transmission function $D(E,\xi)$, while the latter is required 
for the kernels of the embedding equation, from which $Q(E|E^\prime;\eta\eta^\prime)$ 
and subsequently ${\cal B}(E\eta|E^\prime\eta^\prime)$ follow.

Keeping the model flexible and transparent, we use an augmented effective mass model, 
requiring only a few readily accessible material parameters. The simplest model of 
this kind for the surface potential is an image step as shown in 
Fig.~\ref{MethodCartoon}a. To also account for energy gaps in the transmission function 
$D(E,\xi)$, we follow Garc\'ia and Solana~\cite{GS76} and augment the potential on the solid 
side, that is, for  $z<0$ by a potential $V(z)$ periodic in $z$. Using also results from 
MacColl~\cite{MacColl39}, the reflection probability in the two-band approximation is then 
given by  
\begin{align}
R(E,\xi)=\bigg|\frac{\sqrt{\tilde{E}_z}-\sqrt{E_z}\,y+C_\pm[\sqrt{\tilde{E}_z}-\sqrt{E_z}y-G]}
{\sqrt{\tilde{E}_z}+\sqrt{E_z}\,y^*+C_\pm[\sqrt{\tilde{E}_z}+\sqrt{E_z}\,y^*-G]}\bigg|^2
\end{align}
with $E_z=E\xi^2$, $\tilde{E}_z=E_z+U$, and 
\begin{align}
y=-2W^\prime_{-i/4\sqrt{E_z},1/2}(i\sqrt{E_z}/U)/W_{-i/4\sqrt{E_z},1/2}(i\sqrt{E_z}/U)~, 
\end{align}
where $W_{k,\lambda}(z)$ are Whittaker functions and $()^\prime$ denotes the derivative 
with respect to the argument of the function. Using relations between Whittaker functions 
and their derivatives, the ratio $y$ can be expressed as a continued fraction~\cite{LC77} 
and determined numerically. The parameter 
\begin{align}
C_\pm=\frac{E_\pm(G/2)-E_0(G/2)}{V_G}~,
\end{align}
with $E_0(G/2)=(G/2)^2-U$ and $E_\pm(G/2)=E_0(G/2) \pm V(G)$, contains the information about 
the Bragg gap, which forms--within the two-band model--at wave number $k=G/2$. It can be 
verified by inspection that $R(E,\xi)=1$ for $C=\pm 1$, leading to total 
reflection for energies inside the gap. The model for the Bragg gap can be used in two ways. 
Either one calculates $V_G$ from the Fourier transform of the pseudopotential of the ion cores,
setting $G=2\pi n_\mathrm{B}/d$, where $d$ is the spacing between the lattice planes and 
$n_\mathrm{B}$ is the order of the Bragg gap, or one uses it as an effective model, identifying
$E_+(G/2)$ and $E_-(G/2)$ with the experimentally found upper and lower edges of the gap, 
$E_\mathrm{U}$ and $E_\mathrm{L}$, respectively. In that case, $V_G=(E_\mathrm{U}-E_\mathrm{L})/2$.  
In this work we pursue only the second approach.

In the bulk, we include inelastic scattering between electrons and elastic scattering on impurities, 
phonons, and ion cores. Phonon scattering is thus described in the quasi-elastic approximation. The 
scattering on the ion cores is included without an energy threshold, although one would perhaps expect 
it to be operative only at much higher energies~\cite{KB08}. However, even an electron approaching 
the surface with vanishing impact energy has, after traversing the metal's surface potential, an 
energy in the conduction band which is equal to the depth $U$ of the potential. For the metals 
considered, the corresponding de Broglie wave length turns out to be on the order of the lattice 
spacing. Since, in addition, electrons are scattered in arbitrary directions, not only in  
high-symmetry directions, secondary electrons may be affected by the ion cores. Indeed, we found
it essential to include this scattering process to get numerical results in agreement with 
experimental data. The model we use for the bulk is thus essentially the randium model of 
Bauer~\cite{Bauer70}.  

The transition rate resulting from the scattering processes listed above is visualized in
Fig.~\ref{MethodCartoon}c. It is the sum of the Fermi Golden Rule rates for the individual 
processes. Expressing electron momenta in terms of total energies, direction cosines, and 
azimuth angles (spherical coordinates in momentum space with the outward directed surface 
normal as the $z-$axis), and distinguishing between forward and backward scattering, 
depending on the sign of the $z-$components of the electron momenta, yields the expressions 
we now give without calculation.

The rate due to elastic scattering processes becomes
\begin{align}
W_{\rm el}^\pm(E\eta|E^\prime\eta^\prime)&=
16n_{\rm e}\bigg[\big(x_{\rm imp} + 3\frac{k_\mathrm{B}T}{E_\mathrm{F}}\lambda\big)\langle V(g^\pm,0)^2 \rangle_\phi\nonumber\\
&+Z_{\rm v}\langle V(g^\pm,\alpha_c)^2\rangle_\phi\bigg]
\delta(E-E^\prime)
\label{WelInt}
\end{align}
with the momentum transfer 
\begin{align}
g^\pm=|\vec{k}-\vec{k}^\prime|^\pm=g^\pm(E\eta|E^\prime\eta^\prime;\phi)
\label{gfct}
\end{align}
written in terms of total energies, direction cosines, and the difference of the azimuth angles, 
which in Eq.~\eqref{WelInt} is integrated over according to $\langle (...) \rangle_\phi=\int_0^{2\pi}d\phi\,(...)$. 
The parameters $n_e$, $x_{\rm imp}$, $\lambda$, and $Z_{\rm v}$ denote, respectively, the electron density 
of the metal, the impurity concentration in units of $n_e$, the electron-phonon coupling parameter, and 
the valence of the ion cores. The function 
\begin{align}
V(q,\alpha_c)=\frac{1}{q^2 +\kappa^2}\bigg(1-\frac{\alpha_c}{8\pi Z_{\rm v}}\frac{q^2}{(1+(qr_c)^2)^2}\bigg)
\label{Vpseudo}
\end{align}
becomes for $\alpha_c\neq 0$ Harrison's empirical ion pseudopotential~\cite{Harrison65} while for 
$\alpha_c=0$ it reduces to the H\"uckel potential with the screening wave number $\kappa$ of a degenerate 
electron gas at room temperature and density $n_e$.

The rate for electron-electron scattering contains direct and exchange terms. In terms of the variables 
we use, the exchange terms force the expression for the rate to be rather clumsy. For the 
moment, we thus give only the rate due to the direct terms, 
\begin{align}
W^\pm_\mathrm{ee}(E\eta|E^\prime\eta^\prime)|_D=\langle V(g^\pm,0)^2 N(g^\pm,E^\prime-E)\rangle_\phi
\label{WeeD}
\end{align}
with $V(q,\alpha_c)$ given by~\eqref{Vpseudo} and 
\begin{align}
N(q,\omega)=\frac{1}{q\pi^2}\int_{\varepsilon_{\rm min}(q,\omega)}^\infty\!\!\!\!\!\!\!\!\!\!\!\!d\varepsilon\,
n_F(\varepsilon+\omega)[1-n_F(\varepsilon)]~,
\end{align}
where $\varepsilon_{\rm min}(q,\omega)=(\omega-q^2)^2/4q^2$ and $n_F(\varepsilon)$ is the 
Fermi function for the conduction band electrons. The full transition rate, 
$W^\pm_\mathrm{ee}$, including the exchange terms, is however used to produce the data
shown in Section~\ref{Results}. It is given in Appendix~\ref{AppendixC}.

The connection of the transition rates to the kernels of the embedding equation can be found in 
Appendix~\ref{AppendixA}. With the material parameters given in Table~\ref{MaterialParameters} 
we have everything together to calculate the emission yields for the metals listed. Our 
approach is geared towards low impact energies, where solid state effects become important. 
Electron emission caused by high energy electrons is not addressed. 

\section{Results}
\label{Results}

We now present calculated emission yields for different metal surfaces and compare
them to measured data. The material parameters for the metals are 
summarized in Table~\ref{MaterialParameters}. Electron-electron exchange scattering is 
included, although the Monte-Carlo integration it requires increases the computational 
time by two orders of magnitude. Depending on energy and sample, it may change, however, 
the yields up to twenty percent and can thus not be excluded. A few remarks about the 
parameters of the numerics are given in Appendix~\ref{AppendixB}. In the plots shown 
below, energies are always measured from the potential just outside the metal (vacuum level).

\begin{figure}[t]
\includegraphics[width=0.99\linewidth]{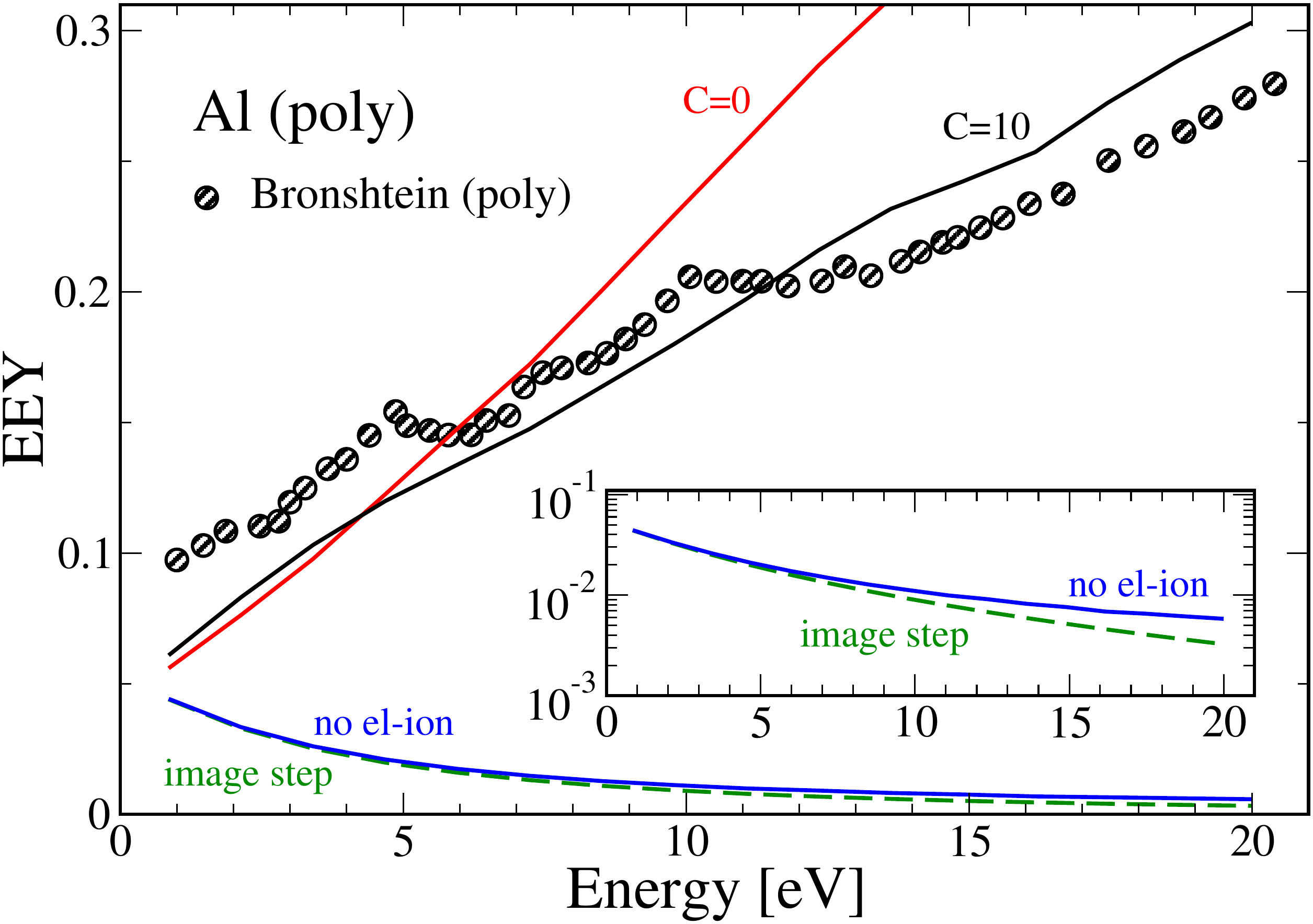}
\caption{(color online)
Emission yield $Y(E,\xi=1)$ due to an electron hitting perpendicularly 
a polycrystalline Al surface with energy $E$. Measured data are from
Bronshtein and Roshchin~\cite{BR58}. Numerical results are shown for an 
idealized perfect surface ($C=0$) and a strongly disordered surface ($C=10$).
For comparison, the yields without incoherent scattering on the ion cores and 
a collisionless image step are also plotted. The inset shows them on a 
log scale for better visibility. 
}
\label{AluminumData}
\end{figure}

We start with the emission yield for aluminum shown in Fig.~\ref{AluminumData}. 
Since the experimental data by Bronshtein and Roshchin~\cite{BR58} (as given by 
Andronov~\cite{Andronov14}) are for polycrystalline aluminum, we do not include Bragg 
scattering on crystal planes parallel to the surface in the calculation. 
We also expect the lateral momentum not to be conserved due to surface defects. 
Indeed, for $C=10$ we obtain better agreement with the measured data than for $C=0$. 
For the ion pseudopotential~\eqref{Vpseudo} we employed a plain H\"uckel potential 
($\alpha_c=0$) because our calculational scheme turned out to be unstable for the 
Al parameters given by Harrison~\cite{Harrison65}.  
Besides the humps around $5\,\mathrm{eV}$ and $10\,\mathrm{eV}$, whose 
origin has to be sought most probably in details of the surface's electronic 
structure beyond the simple two-band model we employ for the calculation of the 
surface transmission function, the theoretical yields are rather close to the 
experimental data. The augmented effective mass model and the approximate calculation 
of the backscattering function $Q$ seem thus to capture the essential physics behind 
secondary emission from polycrystalline Al quite well.

\begin{table}[t]
\begin{center}
  \begin{tabular}{c|c|c|c|c|c|c|c|c|c|c}
 \hline
      & $d [\AA]$ & $\Phi [\mathrm{eV}]$ & $E_{\rm F}[\mathrm{eV}]$ & $\lambda$ & $x_{\rm imp}$ & $Z_{\rm v}$ & $\alpha_c$ & $r_c$ \\\hline\hline
    Au   & 4    & 5.3  & 5.5  & 0.08  & 0.01 & 1 & 83.34 & 0.313 \\
    Ag   & 4    & 4.4  & 5.5  & 0.12  & 0.01 & 1 & 69.0  & 0.457 \\
    Cu   & 3.6  & 4.7  & 7.0  & 0.16  & 0.01 & 1 & 59.1  & 0.516 \\
    Al   & 4    & 4.25 & 11.7 & 0.42  & 0.01 & 3 & 0.0   & --   \\
    W    & 3.16 & 5.22 & 6.4  & 0.28  & 0.01 & 2 & 0.0   & --   \\\hline\hline
  \end{tabular}
  \caption{\small Lattice constant $d$, work function $\Phi$, Fermi energy $E_{\rm F}$,
           electron-phonon coupling constant $\lambda$, impurity concentration $x_{\rm imp}$ 
           in units of the electron density, valence $Z_{\rm v}$, 
           and pseudopotential parameters $\alpha_c$ and $r_c$ from~\cite{Harrison65,Thomas73,BS78}. 
           If not noted otherwise the parameters are in atomic units, with energy measured in
           units of Rydbergs, length in Bohr radii, and mass in bare electron masses.
           The values are typical ones from textbooks~\cite{AM88}. 
  }
  \label{MaterialParameters}
\end{center}
\end{table}

Also shown in Fig.~\ref{AluminumData} is the emission yield obtained by keeping only
electron-impurity and electron-phonon scattering as (quasi-)elastic processes 
competing with the energy loss due to inelastic electron-electron scattering. As can 
be seen, even for an impurity concentration $x_\mathrm{imp}=0.01$, that is, for one 
impurity every 100 electrons, which is rather high, the calculated yield remains way below 
the measured data. It would require a concentration of the order unity, that is, one
impurity for every electron, to bring the theoretical yields up to the measured values.
The only elastic scattering process which brings in such a large factor is the scattering 
of electrons on the potentials of the ion cores. Its rate is proportional to 
$n_{\rm a} Z_{\rm v}^2$, with $n_{\rm a}$ the atomic density of the solid and $Z_{\rm v}$
the valence of the atoms inside the solid. Due to charge neutrality, 
$n_{\rm a}Z_{\rm v}=n_{\rm e}$. Hence, the rate is 
proportional to $n_{\rm e}Z_{\rm v}$. Since $Z_{\rm v}$ is of the order unity, we have a 
process which brings in roughly a scattering center for every electron, as required. 
Without the scattering on the ion cores, the yield is essentially identical to the reflectivity 
of the image step, which is also plotted in Fig.~\ref{AluminumData}. 

For vanishing impact energy the calculated and measured yields do not agree too  
well. The discrepancy could be remedied by increasing the depth of the surface potential. 
From MacColl's calculation~\cite{MacColl39}, however, we infer that the depth 
$U=\Phi+E_{\rm F}$ should be more than $20\,\mathrm{eV}$ (instead of $15.95\, \mathrm{eV}$ 
we use) to push the reflectivity at zero energy up to the experimental value. Without 
surface chemistry, for instance, the formation of an oxide layer, which is beyond the 
scope of the present investigation, it is hard to envisage a solid state effect which 
could increase the potential depth by such a huge amount. Hence, we did not adjust $U$ to 
improve the agreement between experimental and theoretical yields further.

The data plotted in Fig.~\ref{AluminumData} indicate clearly that the scattering 
cascades inside the solid cause secondary emission. This is of course general
wisdom and in fact the basis of all theoretical descriptions, including ours. 
What is surprising, at least to us, is that electron-electron scattering alone 
yields essentially no emission and that incoherent scattering on the ion cores has to
be included. The unexpected necessity to make the ion cores visible to the electrons 
partaking in the scattering cascades arises however from the fact that once inside 
the solid, the electron has a kinetic energy of at least the depth $U$ of the surface 
potential. For Al, $U=15.95\,\mathrm{eV}$, leading to a de Broglie wave length 
$\lambda_\mathrm{dB}=3.1\,\AA$, which is on the order of the lattice spacing $d=4\,\AA$. 
Since, in addition, not only high-symmetry directions are involved in the scattering
cascades, the scattering on the ion cores should play a role in secondary electron
emission, irrespective of the crystallinity of the sample.

\begin{figure}[t]
\begin{minipage}{0.5\linewidth}
\rotatebox{270}{\includegraphics[width=0.8\linewidth]{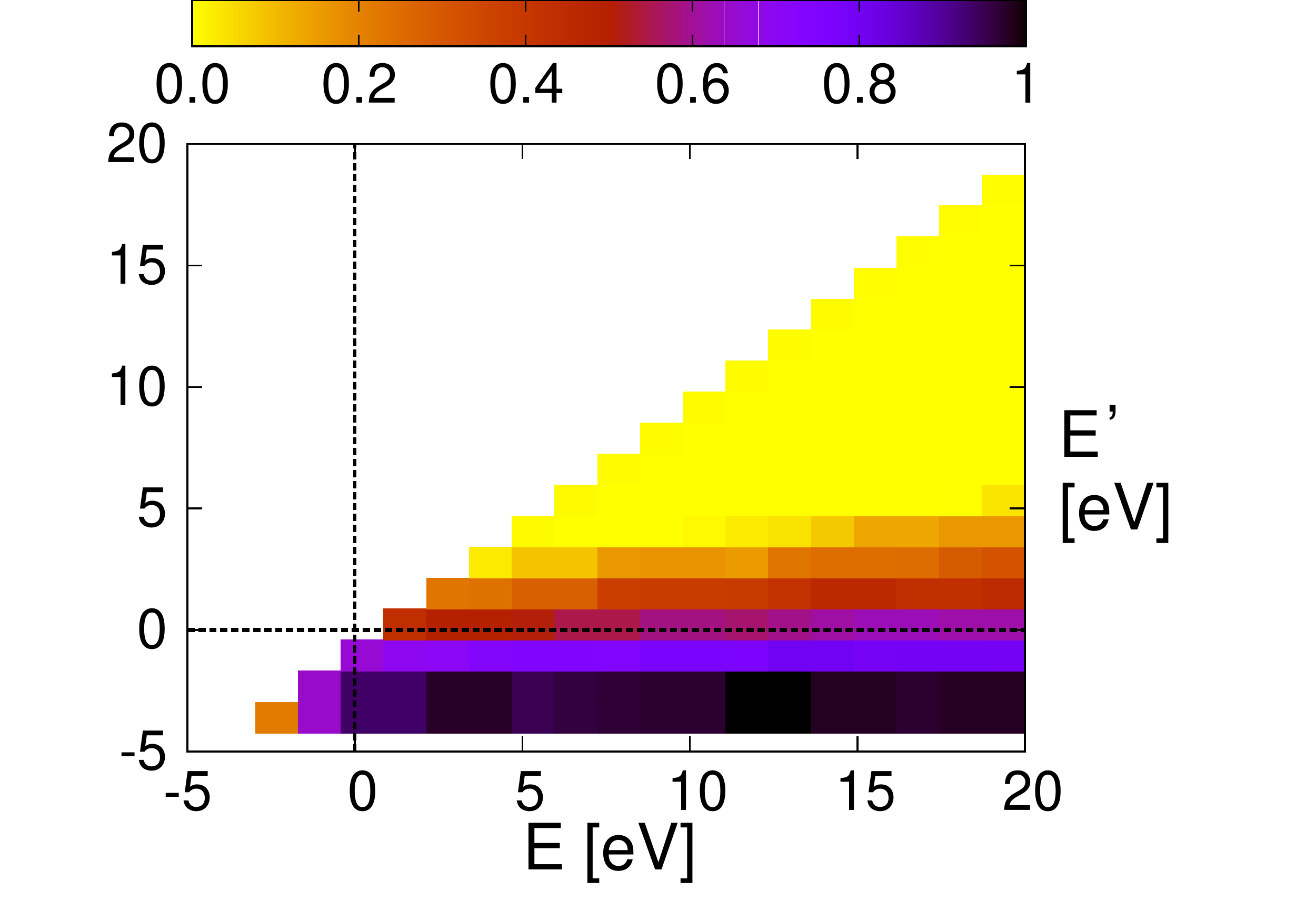}}

\rotatebox{270}{\includegraphics[width=0.8\linewidth]{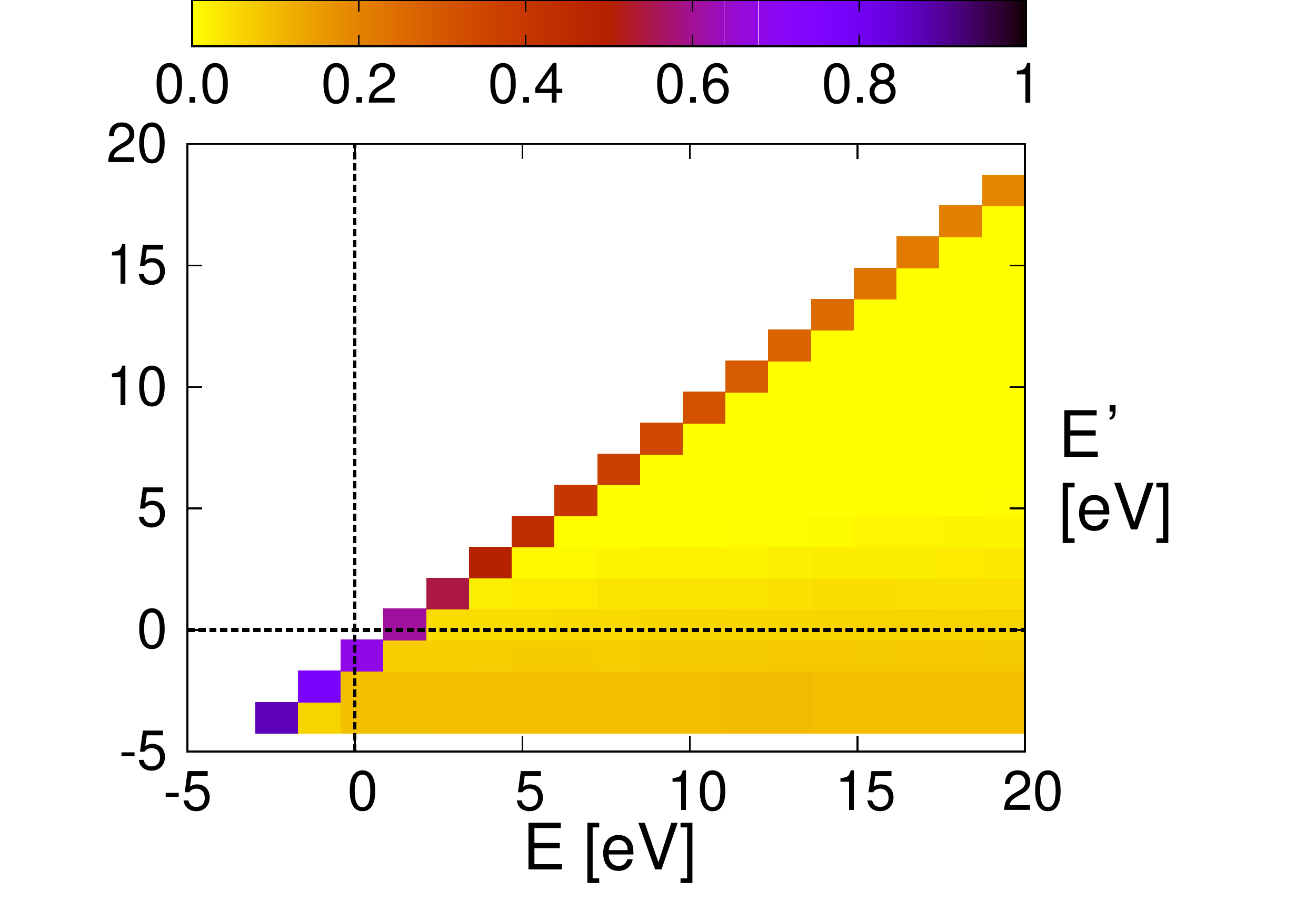}}

\end{minipage}\begin{minipage}{0.5\linewidth}
\rotatebox{270}{\includegraphics[width=0.8\linewidth]{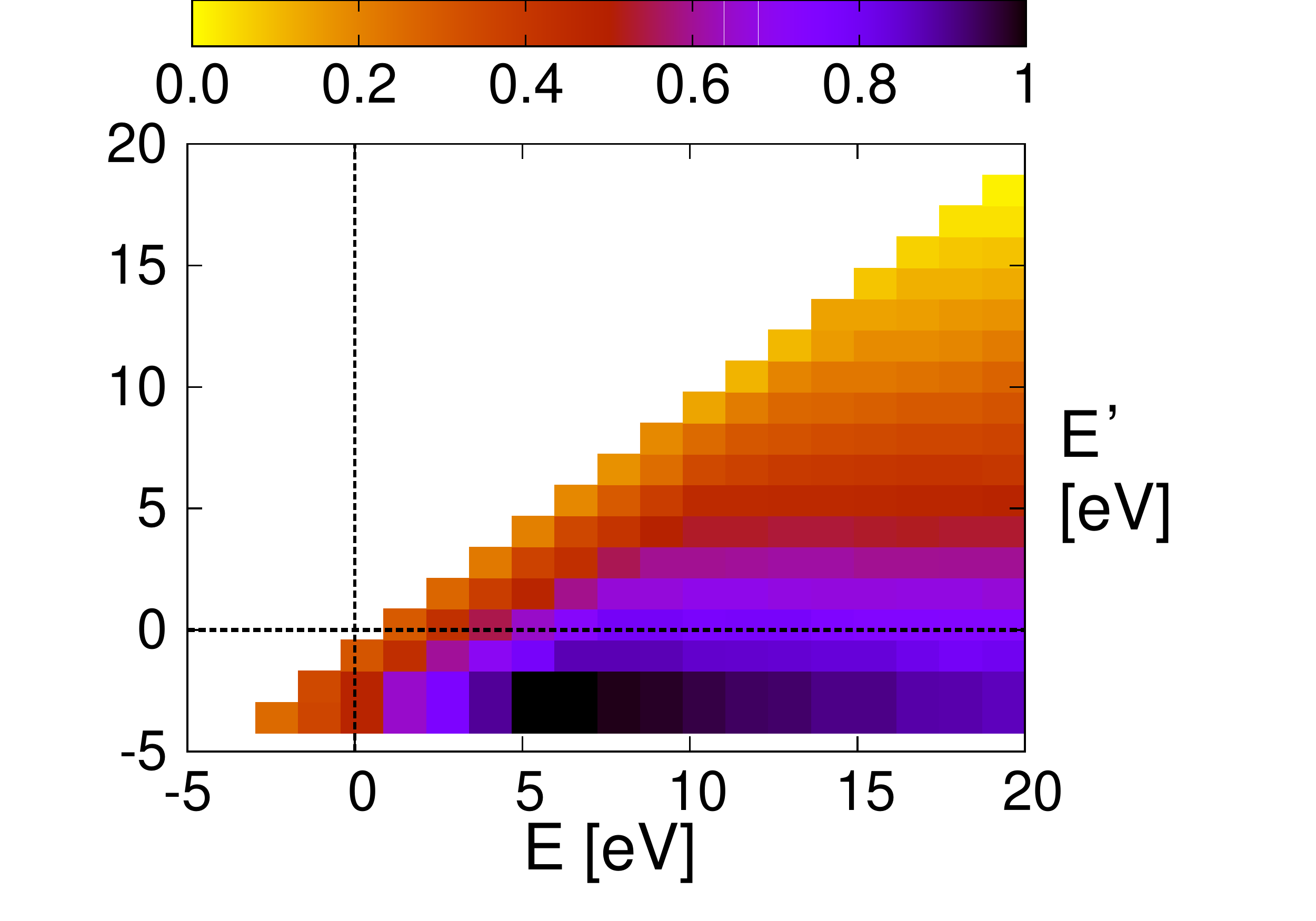}}

\rotatebox{270}{\includegraphics[width=0.8\linewidth]{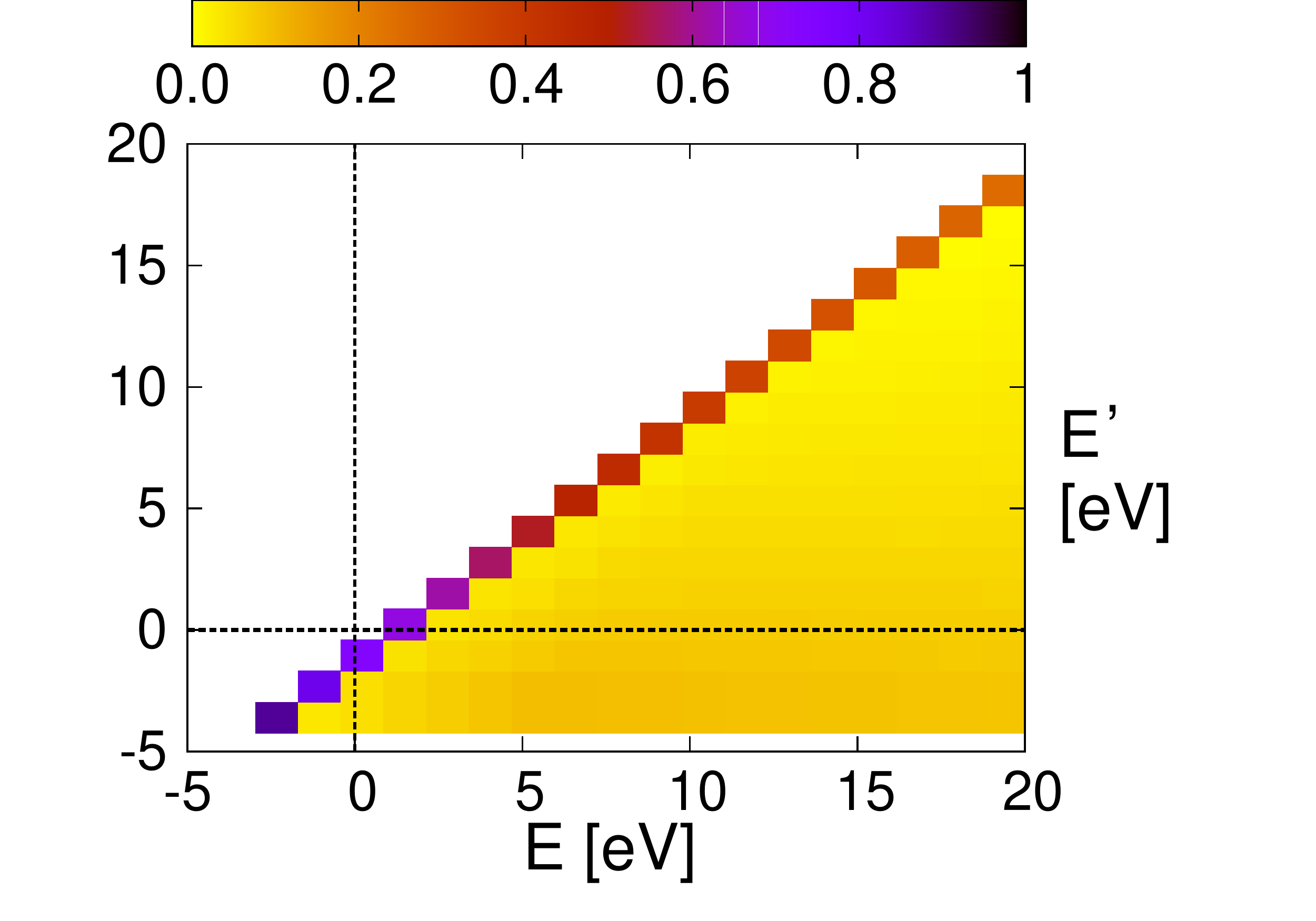}}

\end{minipage}
\caption{(color online) Transition rates for Al using the model described in
Section~\ref{MethodModel}. Left and right panels display
$W^-(E\eta|E^\prime\eta^\prime)$ and $W^+(E\eta|E^\prime\eta^\prime)$
without (top) and with (bottom) incoherent scattering on ion cores included.
The data are normalized to the largest value of the respective transition
rates to indicate the most probable final states more clearly.
The direction cosines are $\eta=1$ and $\eta^\prime=0.3$ (for other
values of $\eta^\prime$, the plots look qualitatively similar) and the
energy resolution $\Delta E\approx 0.7\,\mathrm{eV}$ is the one taken for
the numerical solution of the embedding equation~\eqref{EmbeddingEq} as 
described in Appendix~\ref{AppendixB}.
}
\label{RatesData}
\end{figure}

To understand the vanishing emission yield due to electron-electron scattering, we take
a closer look at the transition rates. Figure~\ref{RatesData} plots on the left and 
right, respectively, $W^-(E\eta|E^\prime\eta^\prime)$ and $W^+(E\eta|E^\prime\eta^\prime)$ 
for $\eta=1$ and $\eta^\prime=0.3$. The top (bottom) row shows the rates without 
(with) the elastic scattering on the ion cores taken into account. To make 
the most probable final states of the scattering event identifiable, 
we normalized the respective rates to their largest values. Clearly, excluding 
the scattering on the ion cores, a backscattered electron very unlikely finds 
itself in a final state above the vacuum level. The most probable states are 
below it, close to the Fermi energy. The same holds for forward scattering. 
Hence, regardless of the scattering direction, due to the strong energy loss 
in an electron-electron collision, electrons in the final states will hardly 
make it over the potential barrier. Including the scattering on the ion 
cores, in contrast, makes elastically backscattered final states above the 
vacuum level accessible, as can be seen in the plots of the bottom row. Due 
to the Volterra-type structure of Eq.~\eqref{EmbeddingEq}, this is most favorable
for electron emission. Hence, a sizeable emission yield can now be expected. 

The scattering on the ion cores makes elastic final states in forward direction
also more likely. However, the gain in backscattering above the vacuum level 
compensates for this detrimental effect. Since the large energy transfer due 
to electron-electron scattering is a general feature of hot electrons in metals,
as can be inferred from the work of Ladst\"adter and coworkers~\cite{LHP04}, 
we suspect, for a secondary electron to get out of a metal, it has to suffer 
along its way inside the solid also incoherent elastic scattering on the ion 
cores, irrespective of the metal's crystallinity. Our results for other metals 
to which we now turn support this conclusion.

\begin{figure}[t]
\includegraphics[width=0.99\linewidth]{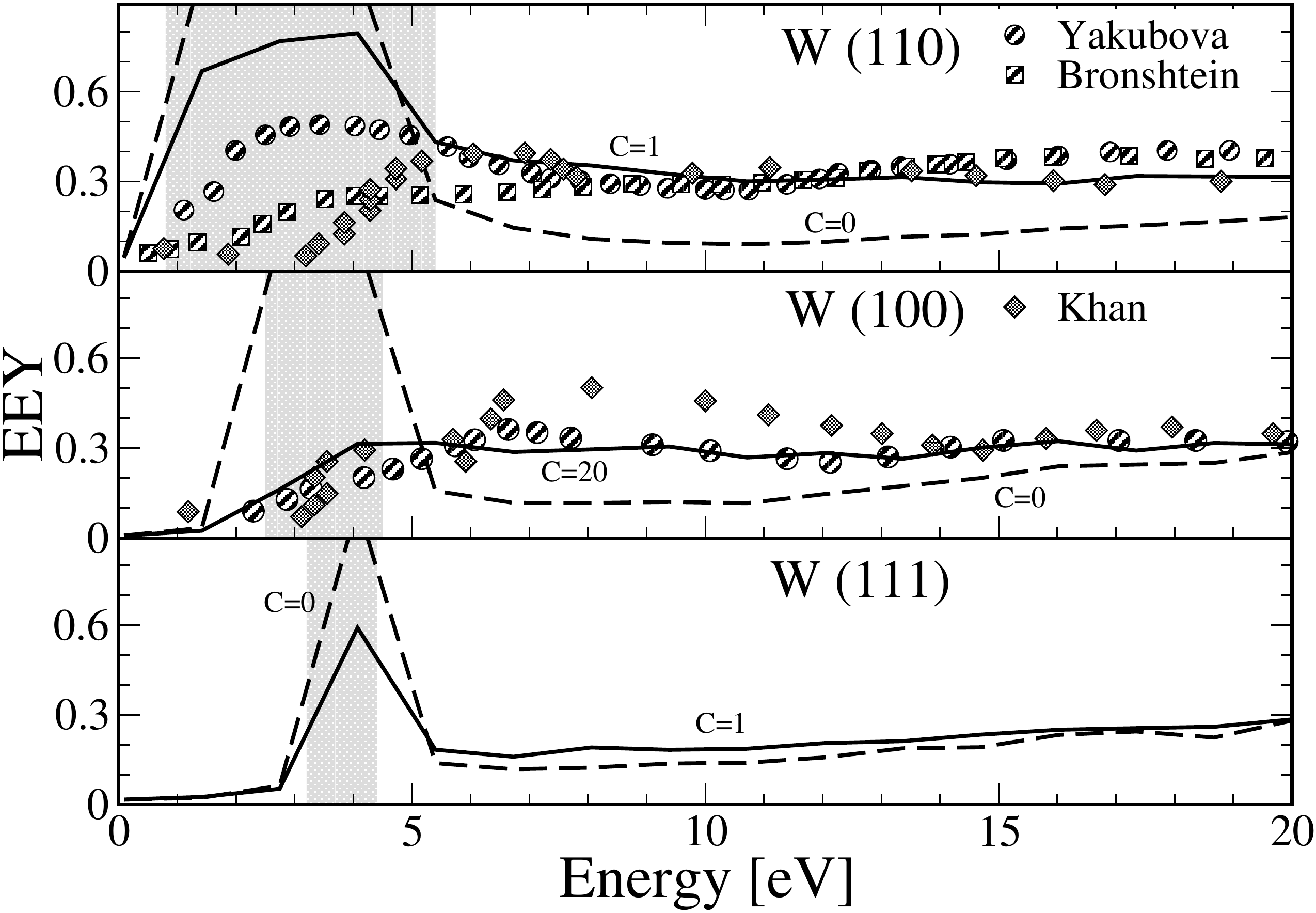}
\caption{
Secondary emission yield $Y(E,\xi=1)$ for different single-crystal tungsten surfaces.
Experimental data are from Khan and coworkers~\cite{KHA63}, Bronshtein and 
Fraiman~\cite{BronFrai69}, and Yakubova and 
Gorbatyi~\cite{YG70,YG72}. Shaded regions indicate Bragg gaps in the electronic structure 
of the surfaces~\cite{Willis75}. The surface parameter $C$ is attached to the theoretical data.
}
\label{TungstenData}
\end{figure}

To test our approach for metal surfaces with a Bragg gap in the energy range 
of interest, we consider tungsten (110), (100), and (111) surfaces. The upper 
and lower edges of the gaps, $E_U=E_+(G/2)$ and $E_L=E_-(G/2)$ in our notation, 
have been determined experimentally by Willis~\cite{Willis75} and can thus be inserted 
into our formalism as described in the previous section. The remaining parameters 
for W are given in Table~\ref{MaterialParameters}. Figure~\ref{TungstenData} plots
the emission yields again for perpendicular impact as a function of energy. The Bragg 
gaps are the regions shaded in grey and measured data are from Yakubova and 
Gorbatyi~\cite{YG70,YG72}, Khan and coworkers~\cite{KHA63}, and Bronshtein and 
Fraiman~\cite{BronFrai69} (as given by Tolias~\cite{Tolias16}). 

Let us start with the data for the W(110) surface, shown in the upper panel of 
Fig~\ref{TungstenData}. For impact energies larger than the energy gap the experimental
data coincide rather well with each other and also with the theoretical yield we obtained 
by setting the surface defect scattering strength $C=1$. In the gap region, 
however, experimental results scatter significantly. Whereas the data from Bronshtein 
and Fraiman~\cite{BronFrai69} as well as Khan and coworker~\cite{KHA63} vanish more or 
less monotonously with decreasing energy, the data from Yakubova and Gorbatyi~\cite{YG70}
show a hump. We interpret the hump to be due reminiscent to the Bragg gap. With the parameter 
$C$ we have control over the emission yield in the gap region. For $C=0$ (dashed line), 
that is, without interfacial scattering, for an ideal surface, the yield in this region 
would be unity. The deviation of the experimental data from unity signals the presence 
of surface defects. Since they couple electron states with different lateral momenta, 
total reflection cannot be maintained over the gap region. With increasing scattering 
strength $C$, the reflectivity reduces thus to less than unity. A reasonable overall fit 
to the data of Yakubova and Gorbatyi~\cite{YG70} we obtained for $C=1$, although in 
the gap region the theoretical yields are still about a factor two too large. 

Emission yields for the (100) and (111) tungsten surfaces are shown, respectively, in the 
middle and lower panels of Fig.~\ref{TungstenData}. For the latter we could not find 
experimental data. The theoretical results are thus predictions. On a finer energy grid,
the gap region would be better resolved, but the expected increase of the emission yield 
can be already seen from the presented data. As far as the (100) surface 
is concerned, the experimental data from Khan and coworkers~\cite{KHA63} and Yakubova and 
Gorbatyi~\cite{YG72} show no hump in the gap region. The surfaces must have been thus rather 
disordered to prevent total reflection. Indeed, setting $C=20$, we reproduce sufficiently 
well the measured yields. The fair agreement between theory and experiment we reach also 
for tungsten is a further indication of the viability of our approach. Let us stress
that our assumption of incoherent scattering on ion cores contributing also in single 
crystals to the scattering cascades responsible for secondary emission could be 
tested by measuring the yields for high-quality, defect-free single-crystal surfaces. If 
the measured data turn out to be below the dashed lines shown in the three panels of 
Fig.~\ref{TungstenData}, scattering on ion cores does not contribute and our assumption 
would be false. 

\begin{figure}[t]
\includegraphics[width=0.99\linewidth]{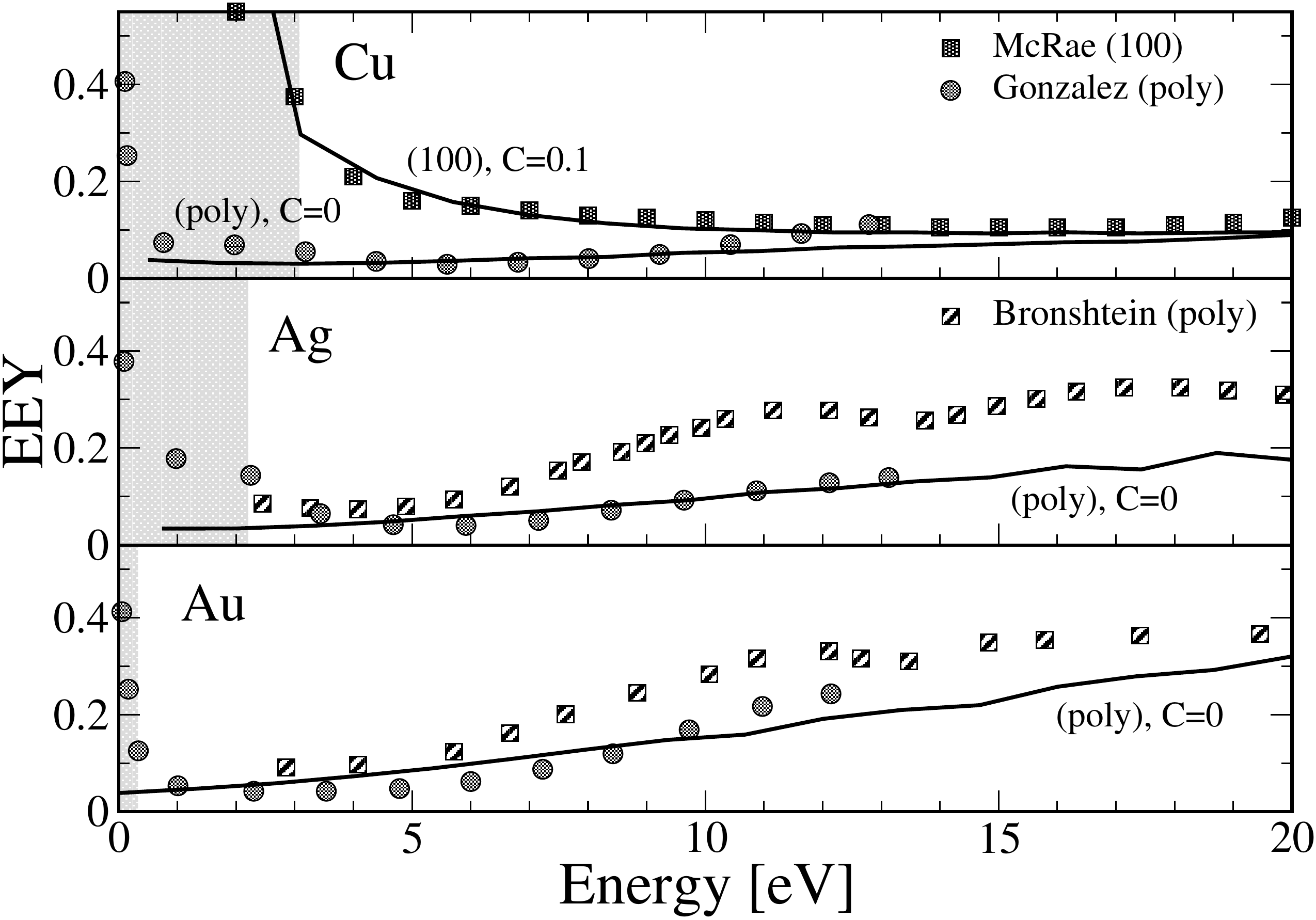}
\caption{
Secondary emission yield $Y(E,\xi=1)$ for different 
noble metal surfaces. Measured data for Cu(100) are from McRae and 
Caldwell~\cite{MC76}, while the data for the polycrystalline Cu, Ag, and Au surfaces are 
from Gonzalez and coworkers~\cite{GAL17} and Bronshtein and Roshchin~\cite{BR58} as 
indicated. The scattering strength $C$ of the surface defects for which good agreement 
between theoretical and measured data is obtained are also given in the plot. For 
reference, the shaded regions indicate the Bragg gaps of the (100) surfaces~\cite{CSE99}.
}
\label{NobleMetalData}
\end{figure}

A comparison of calculated yields for the noble metals Cu, Ag, and Au and measured data
by Gonzalez and coworkers~\cite{GAL17} as well as Bronshtein and Roshchin~\cite{BR58}
(as given by Andronov~\cite{Andronov14}) is shown in Fig.~\ref{NobleMetalData}. 
Avoiding a discussion of chemical modifications
of the as-received surfaces, which are outside the scope of our approach, we focus on 
what Gonzalez and coworkers call clean samples. For Cu, we also include data from McRae and
Caldwell~\cite{MC76} because they were taken for a single crystal (100) surface, showing
a Bragg gap, whereas the other data are for polycrystalline samples, where no Bragg gaps
are expected. The grey regions in the three panels indicate, for purpose of reference, 
the gaps for Cu(100), Ag(100), and Au(100) as obtained from the calculations of 
Chulkov and coworkers~\cite{CSE99}.

In the upper panel we see for Cu(100) almost perfect agreement between experiment and 
theory by setting $C=0.1$. The surface used by McRae and Caldwell~\cite{MC76} must have 
been thus nearly void of defects. As far as the polycrystalline Cu employed by Gonzalez and 
coworkers~\cite{GAL17} is concerned, we get fair agreement by excluding the possibility 
of a Bragg gap and setting $C=0$, implying a clean surface. Since the surface was indeed 
carefully cleaned by sputtering with $\mathrm{Ar}^+$ ions, $C=0$ is perhaps expected. 
The same we find for the polycrystalline, sputter-cleaned Ag and Au surfaces. As can be 
seen in the middle and lower panels of Fig.~\ref{NobleMetalData}, fair agreement between 
measured and calculated data is again obtained for $C=0$, although not as good as for Cu.

We could not reproduce Bronshtein and Roshchin's data~\cite{BR58} for Ag and Au, 
which are for all energies roughly a factor two above the ones measured by Gonzalez and 
coworkers~\cite{GAL17}. The origin of the enhancement remains unclear. Perhaps the 
surfaces were contaminated by ad-atoms or oxide layers, but this is outside the scope 
of the present investigation. Within our model, we have also no explanation for the humps
around $11\,\mathrm{eV}$, which in fact do also not show up in the recent data from  
Gonzalez and coworkers~\cite{GAL17}. Since we do not expect plasmon effects to affect 
backscattering in this energy range, the cause of the humps should be also sought 
in a chemical modification of the surfaces. Further studies, ideally 
based on the solution of the embedding equation which keeps the angular dependence of 
the backscattering function fully intact, are required to clarify the origin of the humps.  

At this point, it is appropriate to comment on the preliminary results for Ag we 
included in a perspective paper about the electron microphysics at plasma-solid 
interfaces~\cite{BRF20}. There, we claimed to reach good agreement between experimental 
and theoretical data by adjusting an energy-dependent electron-phonon coupling strength. 
Incoherent scattering on the ion cores was not considered, as was the scattering on 
surface defects. In addition, the surface potential was the exponential barrier employed 
by Roupie and coworkers~\cite{RJT13}. The studies on which the present report is based 
revealed, however, that the exponential barrier is a bad approximation to the image step 
(irrespective of the numerical value of the additional adjustable parameter, which it brings in) 
and that the good adjustment disappears by incorporating the angular dependence of electron-phonon 
scattering. Since we now find fair agreement for a variety of metals, we consider the
conclusions reached in this work as robust. 

Finally, let us comment on the debate in the plasma physics community~\cite{ASK13,Andronov14}
which arose around the initial measurements by Cimino and coworkers~\cite{CCF04}. In particular, 
their claim that the yields reach unity for vanishing impact energy was critically seen. In 
the meantime, however, Cimino's group scrutinized the data very carefully~\cite{CGL15,GAL17};
the work by Gonzalez and coworkers~\cite{GAL17} is one of the follow-up investigations. 
Specifically in Ref.~\cite{CGL15} they estimated the errors of their measurements at low energies, 
with the conclusion that an unanimous claim for unit reflectivity at zero energy cannot be 
maintained. General arguments, and also the explicit calculation of MacColl~\cite{MacColl39}, 
indeed show that unit reflectivity cannot occur on an image step. A Bragg gap, however, as we 
have seen above for the single-crystal Cu(100) surface, could lead to an increase of the 
reflectivity, and hence, the emission yield. We would in fact expect the same for 
single-crystal Au(100) and Ag(100) surfaces, since they have Bragg gaps around the vacuum 
level~\cite{CSE99} (shaded regions in the middle and lower panels of Fig.~\ref{NobleMetalData}),
and would encourage measurements on them. For the polycrystalline Ag and Au samples used by
Cimino's group, however, this cannot occur, supporting thus Andronov and coworkers'~\cite{ASK13,Andronov14} 
critique of Cimino and coworkers' initial interpretation of the measured data. An explanation 
of the discrepancies between the recent~\cite{GAL17} and the previous~\cite{BR58} noble metal 
data remains however outside our approach. We cannot account for it by the surface scattering 
strength $C$. Most probably, its origin has to be sought in chemical modifications of the 
surface.

\section{Conclusion}
\label{Conclusions}
We presented an equation for the secondary electron emission yield $Y(E,\xi)$, which 
combines the three stages of the process: transmission of a primary electron through the 
surface potential, scattering cascades inside the bulk of the solid exciting secondary 
electrons, and escape of secondaries by transmission through the surface potential 
in the reverse direction. The structure of the expression, resembling a conditional 
(pseudo-)probability in its central part, which is the escape function 
${\cal E}(E,\xi)$, reflects the three stages in a rather transparent manner.  
Due to the explicit consideration of the transmission through the surface potential, 
the expression for $Y(E,\xi)$ is particularly suitable for describing secondary emission 
from surfaces at very low electron impact energies. 

The two main building blocks of the formula for the escape function are the surface 
transmission function $D(E,\xi)$ and the backscattering function $Q(E\eta|E^\prime\eta^\prime)$. 
The former, to be obtained from a quantum-mechanical calculation matching the states inside 
the solid to the ones outside of it, reflects the electronic structure of the surface, while 
the latter, obtained from the invariant embedding principle, contains the bulk scattering 
processes. To account for electron multiplication due to electron-electron scattering, the 
part of the backscattering function arising from it is normalized to two.

In this work, we calculated the surface transmission function for an image step, augmented 
by diffraction on crystal planes parallel to the surface and elastic scattering on surface 
defects. The backscattering function was obtained from an augmented effective mass model, 
including statically screened electron-electron scattering, electron-impurity scattering, 
quasi-elastic electron-phonon scattering, and incoherent scattering on the potentials of 
the ion cores, which turned out to be rather important for getting a good match of calculated 
and measured yields. As a first step of solving the nonlinear embedding equation for 
$Q(E\eta|E^\prime\eta^\prime)$, we decoupled angular and energy integrations by the 
quasi-isotropic approximation. 

Despite the simplicity of the model and the crudeness of the quasi-isotropic approximation, 
calculated and measured data are rather close to each other, indicating the calculational 
scheme captures the essentials of secondary emission from metal surfaces at very 
low electron impact energies. The discrepancies between theory and measurements which remain 
could be due to shortcomings of the quasi-isotropic approximation, which assumes isotropy
of electron backscattering irrespective of the energy transfer, while it is actually only 
the case for elastic processes, or the limitations of the effective two-band model used to 
obtain the surface transmission function $D(E,\xi)$. In principle, the lack of dynamic 
screening and the lack of the bulk crystal structure in the transition rates could be further
causes. But we do not expect it to be critical. From our perspective, the next step should 
be to calculate $Q(E\eta|E^\prime\eta^\prime)$ without the quasi-isotropic approximation. 
Combined with the expression for the emission yield we have given in this work, one would 
then have a powerful alternative to Monte-Carlo simulations of secondary electron emission 
at low impact energies. 

\section*{Acknowledgments}
We thank Kristopher Rasek from the University Greifswald for critical reading of the 
manuscript. F\,.X\,.B. thanks Igor Kaganovich from the Princeton Plasma Physics Laboratory 
for discussions and help with the Russian literature.

\appendix

\section{Quasi-isotropic approximation}
\label{AppendixA}

The goal of the quasi-isotropic approximation is to decouple the energy from the angular 
variables and to reduce thereby the embedding equation, which is a four-dimensional 
integral equation, to a two-dimensional one. 

The embedding equation, as it arises from the invariant embedding 
principle for electron backscattering~\cite{Dashen64}, can be written as  
\begin{align}
Q&=K^- + K_1^+ \circ Q + Q \circ K_2^+ + Q \circ B^- \circ Q
\label{FullEmbeddingEq}
\end{align}
where the $\circ$ operation is defined by 
\begin{align}
(A \circ B)(E\eta|E^\prime\eta^\prime)&=\int_{E^\prime}^E\!\!\!dE^{\prime\prime}
\int_{\eta_c}^1 \!\!\!d\eta^{\prime\prime}
A(E\eta|E^{\prime\prime}\eta^{\prime\prime})
\nonumber\\
&\times B(E^{\prime\prime}\eta^{\prime\prime}|E^\prime\eta^\prime)~,
\end{align}
and a cut-off $\eta_c=10^{-4}$ is introduced to exclude extreme grazing entrance and exit 
angles, which are hardly realized experimentally, but would require the handling of integrable 
singularities, making the numerics thus unnecessarily involved. 

The kernels of~\eqref{FullEmbeddingEq} are given by 
\begin{align}
K^-(E\eta|E^\prime\eta^\prime)&=\frac{G^-(E\eta|E^\prime\eta^\prime)}{H(E\eta|E^\prime\eta^\prime)}~,
\label{Km}
\\
K_1^+(E\eta|E^{\prime\prime}\eta^{\prime\prime};E^\prime\eta^\prime)&=
\frac{G^+(E\eta|E^{\prime\prime}\eta^{\prime\prime})\rho(E^{\prime\prime})}{H(E\eta|E^\prime\eta^\prime)}~,
\\
K_2^+(E^{\prime\prime}\eta^{\prime\prime}|E^\prime\eta^\prime;E\eta)&=
\frac{\rho(E^{\prime\prime})G^+(E^{\prime\prime}\eta^{\prime\prime}|E^\prime\eta^\prime)}
{H(E\eta|E^\prime\eta^\prime)}~,
\\
B^-(E^{\prime\prime}\eta^{\prime\prime}|E^{\prime\prime\prime}\eta^{\prime\prime\prime};E\eta E^\prime\eta^\prime)&=
\nonumber\\
&\frac{\rho(E^{\prime\prime})G^-(E^{\prime\prime}\eta^{\prime\prime}|E^{\prime\prime\prime}\eta^{\prime\prime\prime})\rho(E^{\prime\prime\prime})}
{H(E\eta|E^\prime\eta^\prime)}
\end{align}
with 
\begin{align}
G^\pm(E\eta|E^\prime\eta^\prime)=\sum_{i=\mathrm{ee,el}}\frac{W_i^\pm(E\eta|E^\prime\eta^\prime)}{v(E)\eta}
\end{align}
the transition rate per length, obtained from the individual transition rates per time 
$W_i^\pm(E\eta|E^\prime\eta^\prime)$, where $i=\mathrm{ee}, \mathrm{ei}$, presented in 
Section~\ref{MethodModel}. The function   
\begin{align}
H(E\eta|E^\prime\eta^\prime)=\frac{\gamma(E)}{v(E)\eta}+\frac{\gamma(E^\prime)}{v(E^\prime)\eta^\prime}~,
\end{align}
usually written on the left side of~\eqref{FullEmbeddingEq}, denotes the total rate per length that the
impinging electron scatters in the infinitesimally thin additional material layer envisaged in the embedding 
principle. It is given in terms of the electron velocity $v(E)=2\sqrt{E+U}$ and the total scattering
rate per time, 
\begin{align}
\gamma(E)&=\sum_{i=\mathrm{ee,el}} C_i \int_{-\Phi}^E \!\! dE^{\prime} \int_{0}^1 \!\! d\eta^{\prime}
\rho(E^{\prime})
\bigg[W_i^+(E\eta|E^{\prime}\eta^{\prime}) 
\nonumber\\
&+ W_i^-(E\eta|E^{\prime}\eta^{\prime})\bigg]~,
\label{GammaFct}
\end{align}
which in fact is independent of $\eta$. The factor $C_{\rm ee}=1/2$ avoids double counting of the 
final states in an electron-electron scattering event~\cite{PAG85} and $C_{\rm el}=1$. 

The reasoning behind the quasi-isotropic approximation is that backward scattering, encoded 
in $K^-$, depends only weakly on the angular variables. It simplifies the embedding 
equation~\eqref{FullEmbeddingEq} significantly, since it provides the freedom to fix on 
the rhs, under the integrals over the direction cosines, the angular dependencies of the 
backscattering functions to the ones of the backscattering function on the lhs. This opens 
up the possibility to solve~\eqref{EmbeddingEq} iteratively by an expansion in the number of 
backscattering events (all with the same values for $\eta$ and $\eta^\prime$) which encounters 
in each iteration step just a linear integral equation in two energy variables.

Before discussing how valid the quasi-isotropic approximation in fact is, let us state the 
kernels to which it leads. Utilizing it turns $\eta$ and $\eta^\prime$ to external variables. 
Equation~\eqref{FullEmbeddingEq} reduces then to~\eqref{EmbeddingEq} given in 
Section~\ref{MethodBackScattSect} with
\begin{align}
K_1^+(E|E^{\prime\prime};E^\prime\eta\eta^\prime)&=\int_{\eta_c}^1 \!\!d\eta^{\prime\prime} 
                       K_1^+(E\eta|E^{\prime\prime}\eta^{\prime\prime};E^\prime\eta^\prime)~,\label{K1p}\\
K_2^+(E^{\prime\prime}|E^\prime;E\eta\eta^\prime)&=\int_{\eta_c}^1 \!\!d\eta^{\prime\prime}
K_2^+(E^{\prime\prime}\eta^{\prime\prime}|E^\prime\eta^\prime;E\eta)~,
\end{align}
and 
\begin{align}
B^-(E^{\prime\prime}|E^{\prime\prime\prime};E\eta E^\prime\eta^\prime)&=\nonumber\\
\int_{\eta_c}^1 \!\! d\eta^{\prime\prime} \!\!\!\int_{\eta_c}^1 \!\! d\eta^{\prime\prime\prime}&
B^-(E^{\prime\prime}\eta^{\prime\prime}|E^{\prime\prime\prime}\eta^{\prime\prime\prime};E\eta E^\prime\eta^\prime)~.
\label{Bm}
\end{align}

Let us now address the validity of the quasi-isotropic approximation. The angular dependence 
of $K^-$ decides whether it is justified or not. To be valid, it should be nearly isotropic 
because, in leading order, Eq.~\eqref{FullEmbeddingEq} suggests $Q=K^-$. Hence, provided 
$K^-$ has a weak angular dependence, so will $Q$. By anticipating this, we can thus move 
in the three integral terms of~\eqref{FullEmbeddingEq} the backscattering functions $Q$ in 
front of the angular integrals and set the direction cosines of the $Q$'s to $\eta$ and 
$\eta^\prime$, in the manner indicated above, that is, to the direction cosines on which 
the inhomogeneity $K^-$ and the $Q$ on the lhs depend. 

\begin{figure}[t]
\begin{minipage}{0.5\linewidth}
\rotatebox{270}{\includegraphics[width=0.8\linewidth]{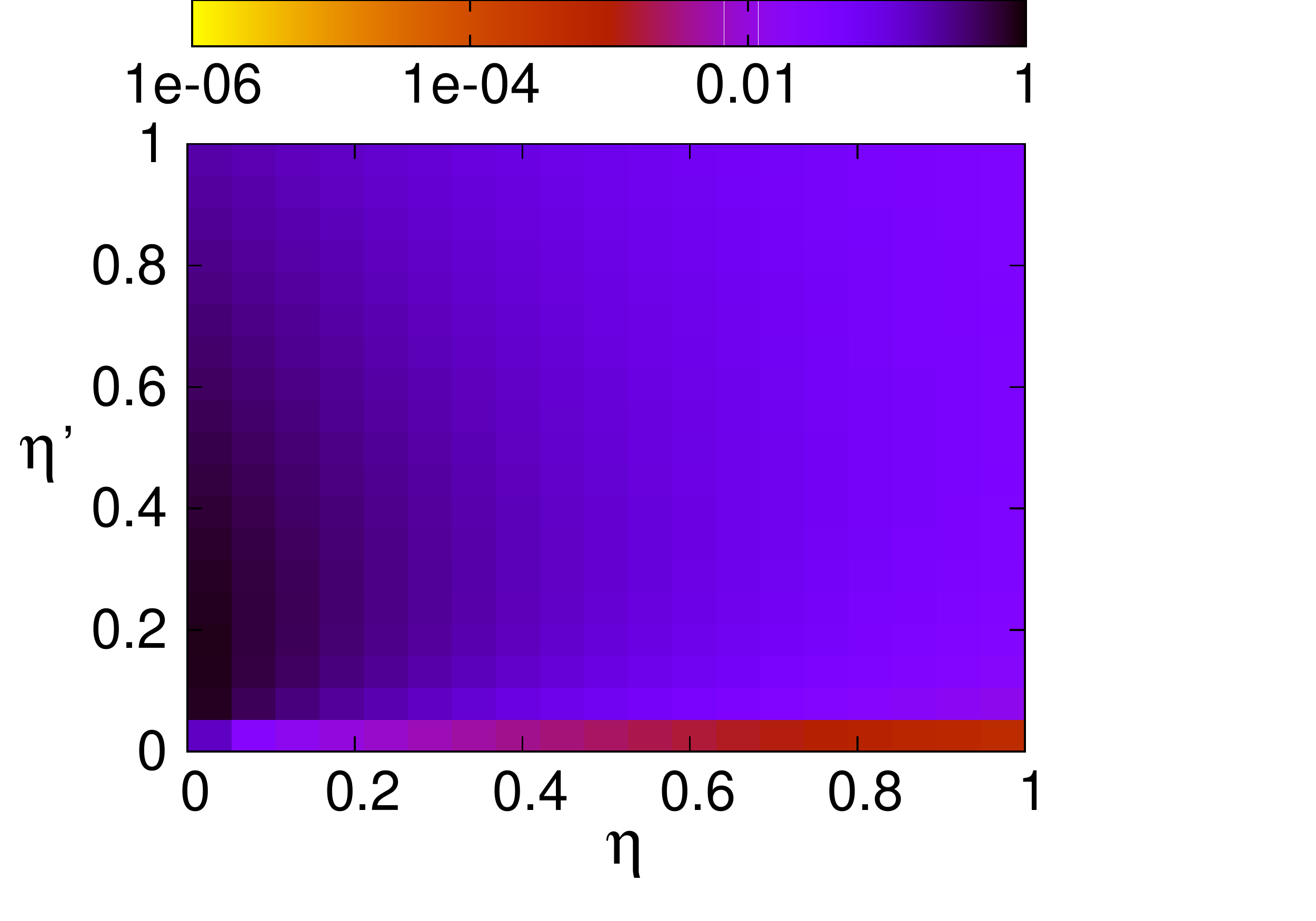}}

\rotatebox{270}{\includegraphics[width=0.8\linewidth]{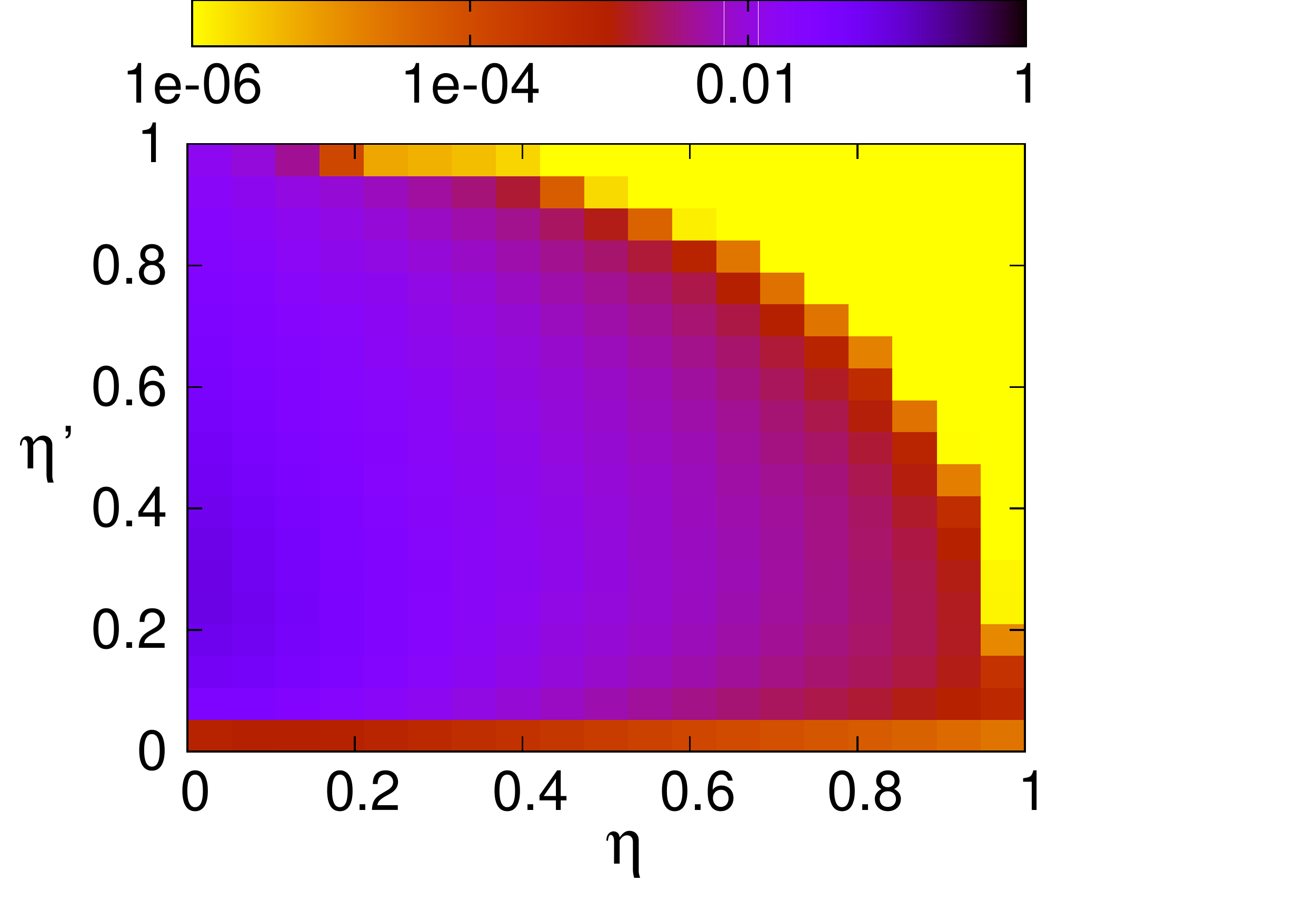}}

\end{minipage}\begin{minipage}{0.5\linewidth}
\rotatebox{270}{\includegraphics[width=0.8\linewidth]{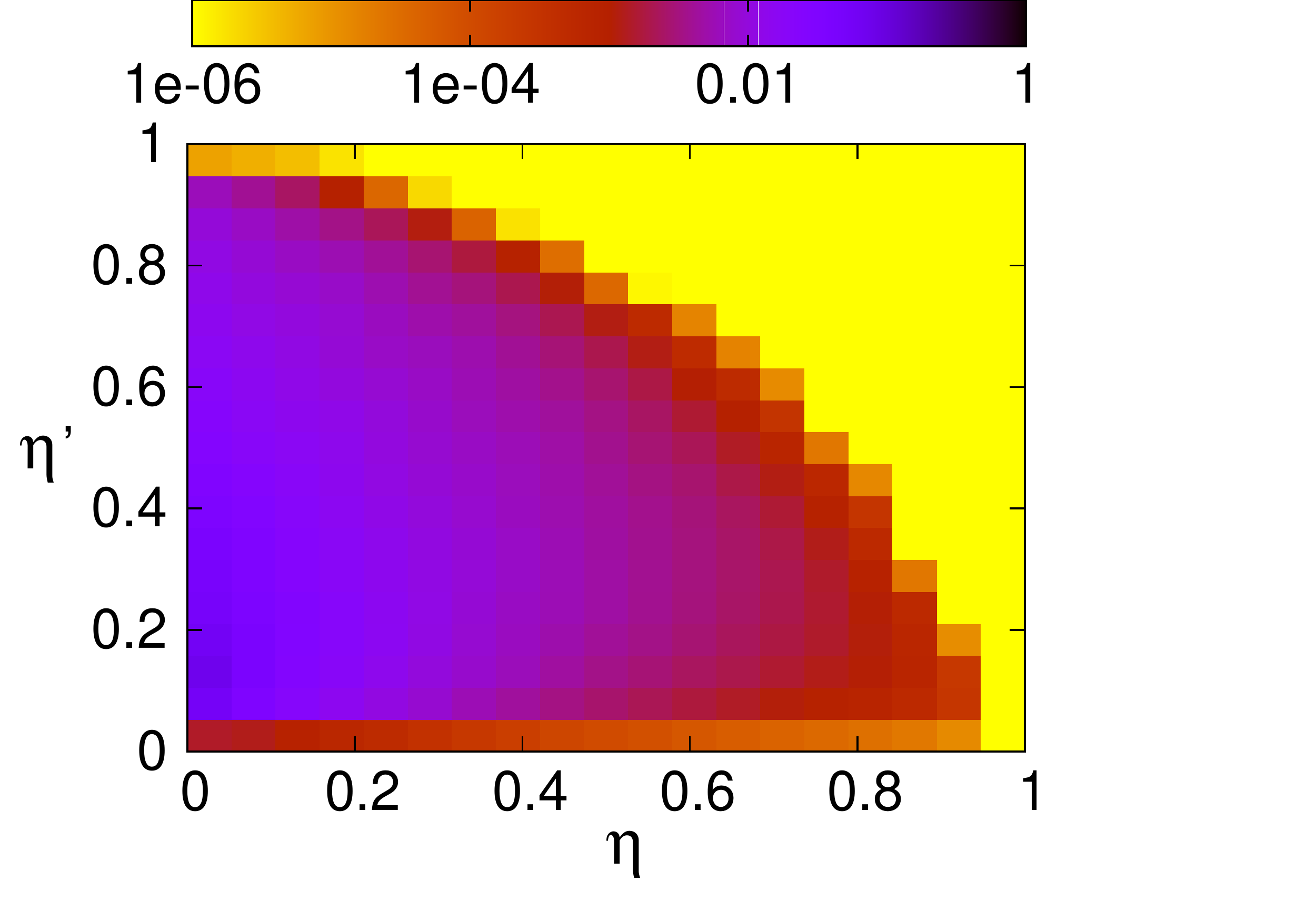}}

\rotatebox{270}{\includegraphics[width=0.8\linewidth]{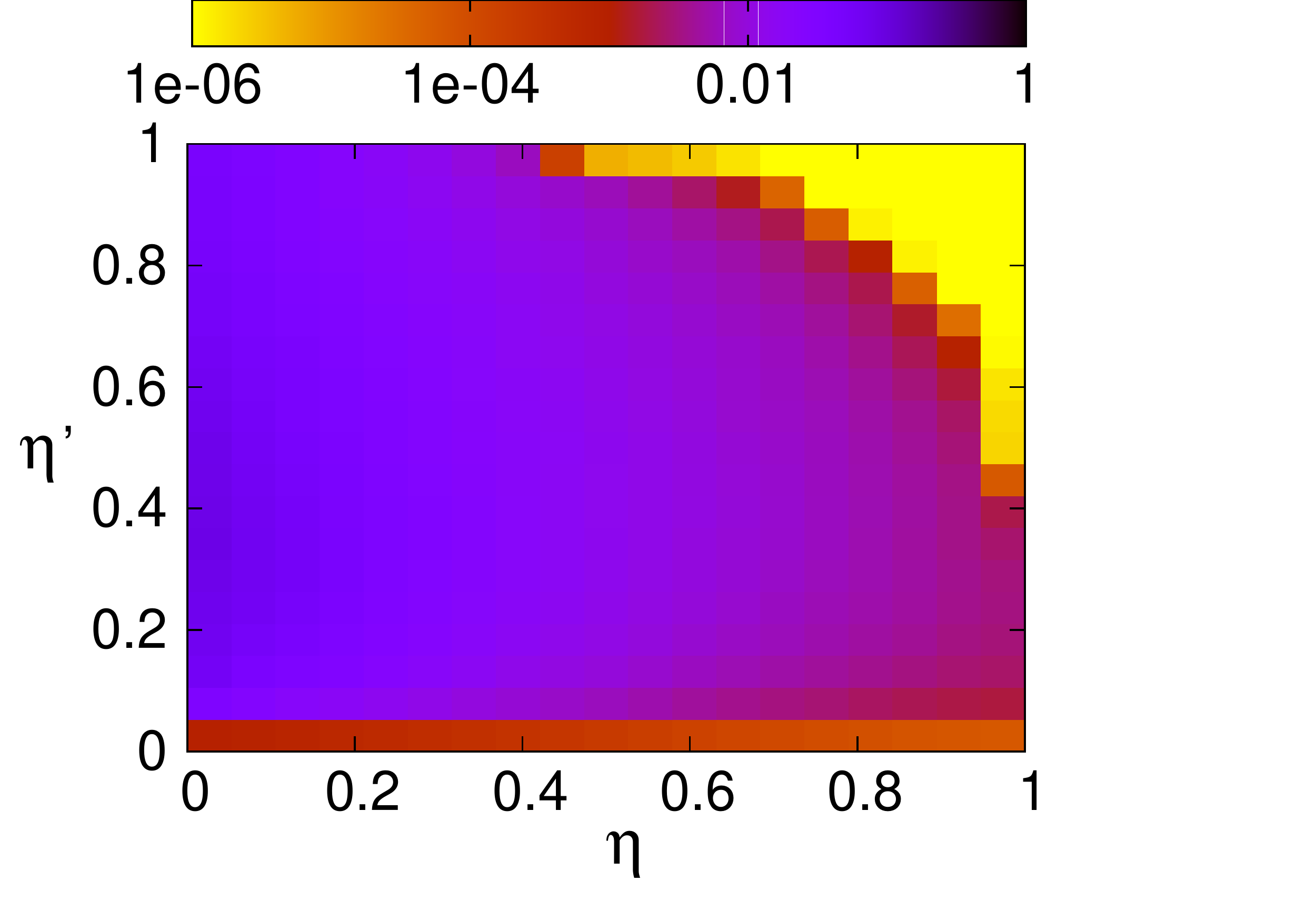}}

\end{minipage}
\caption{(color online)
Angular dependence of $K^-(E\eta|E^\prime\eta^\prime)$ for
Al using the model described in Section~\ref{MethodModel}. From the upper left
moving clockwise through the plots, the initial energy $E=17.45\,\mathrm{eV}$,
while the final energies are $E^\prime=17.45$, $12.34$, $7.24$, and
$2.13\,\mathrm{eV}$. The nonmonotonous growth of the Pauli-blocked zone
with energy transfer (yellow regions) is due to the denominator
$\eta v(E)H(E\eta|E^\prime\eta^\prime)$ in the definition~\eqref{Km} of $K^-$.
Angular resolution is the one taken for the numerical
solution of Eq.~\eqref{EmbeddingEq} sketched in Appendix~\ref{AppendixB}.
}
\label{TransitionRates}
\end{figure}

A representative angular dependence of $K^-(E\eta|E^\prime\eta^\prime)$ is plotted in
Fig.~\ref{TransitionRates}, using the model described in Section~\ref{MethodModel} and 
material parameters for Al. Starting with elastic scattering in the upper left and 
moving clockwise through the plots, data are shown for increasing energy transfer. 
Clearly, elastic backscattering is rather isotropic. The quasi-isotropic approximation
is well justified for it. For finite energy transfer, however, that is,
for inelastic backscattering due to electron-electron collisions, isotropy is no longer
strictly given, especially, for large direction cosines, where the Pauli principle 
excludes final states. The quasi-isotropic approximation neglects this and assumes, 
irrespective of the energy transfer, $K^-(E\eta|E^\prime\eta^\prime)$ to be a weakly 
varying function of its angular variables. Since the reduction of computational costs 
is large and the agreement between calculated and measured yields is rather good, we 
consider the approximation as a viable first step towards a complete solution of the 
embedding equation, which then, of course, has to keep energy and angular variables 
fully intact.

\section{Solution of Eq.~\eqref{EmbeddingEq}}
\label{AppendixB}

In this appendix, we describe the strategy we employed for solving the embedding equation in the 
quasi-isotropic approximation. To grasp the algebraic structure of the approach, we suppress in 
the following the independent variables and adopt a symbolic notation, in which the 
separation~\eqref{Splitting} simply reads
\begin{align}
A=A_{\rm ee}+A_{\rm el}\delta(E-E^\prime)~.
\end{align}

Using the splitting for the functions entering~\eqref{EmbeddingEq}, collecting 
all terms proportional to $\delta(E-E^\prime)$, and forcing them to vanish yields 
an algebraic equation for $Q_{\rm el}$ and an integral equation for $Q_{\rm ee}$.
The algebraic structure of the two equations is identical. In an abstract notation, 
they both can be cast into the form 
\begin{align}
Q_{\rm i}=K^-_{\rm i} + K^+_{\rm 1,i} Q_{\rm i} + Q_{\rm i} K^+_{\rm 2,i} + Q_{\rm i} B_i^- Q_{\rm i}~.
\end{align}
with ${\rm i=el, ee}$. For ${\rm i=el}$ the kernels are just the factors in $K^-$ 
and $K^+_{1,2}$ which are in front of $\delta(E-E^\prime)$, while for ${\rm i=ee}$
the kernels are renormalized and given by 
\begin{align}
K^-_{\rm ee} &\rightarrow \tilde{K}^-_{\rm ee}=\frac{N^-}{D}~,~~ 
K^+_{1,{\rm ee}} \rightarrow \tilde{K}^+_{1,{\rm ee}}=\frac{N^+_1}{D}~,\\ 
K^-_{2,{\rm ee}} &\rightarrow \tilde{K}^+_{2,{\rm ee}}=\frac{N^+_2}{D}~,~~
B^-_{\rm ee} \rightarrow \tilde{B}^-_{\rm ee}=\frac{B^-_{\rm ee}}{D}
\end{align}
with 
\begin{align}
D &= 1-K^+_{1, {\rm el}} - K^+_{2, {\rm el}} - Q_{\rm el} B^-_{\rm el} - B^-_{\rm el} Q_{\rm el}~,
\\
N^- &= K^-_{\rm ee} + Q_{\rm el} B^-_{\rm ee} Q_{\rm el} + K^+_{\rm 1, ee} Q_{\rm el}~
       + Q_{\rm ee} K^+_{\rm 2, ee}~, 
\\
N^+_1 &=K^+_{\rm 1, ee} + Q_{\rm el} B^-_{\rm ee}~,
\\
N^+_2 &=K^+_{\rm 2, ee} + B^-_{\rm ee} Q_{\rm el}
\end{align}
and the energy variables of the (one-energy) functions $Q_{\rm el}$ such that they coincide,
depending on the relative ordering, with the left or right energy variables of the (two-energy) 
functions next to them.  

The algebraic equation for $Q_{\rm el}$ is readily solved. Defining an auxiliary function 
\begin{align}
Q^{(1)}_{\rm el}=\frac{K^-_{\rm el}}{1-K^+_{\rm 1,el}-K^+_{\rm 2,el}}
\end{align}
one obtains 
\begin{align}
Q_{\rm el}=\frac{K^-_{\rm el}}{2B^-_{\rm el}Q^{(1)}_{\rm el}}\bigg[ 1-
\sqrt{1-\frac{4B^-_{\rm el}\big(Q^{(1)}_{\rm el}\big)^2}{K^-_{\rm el}}} \bigg]~,
\label{Static}
\end{align}
provided $D=1-4B^-_{\rm el} \big(Q^{(1)}_{\rm el}\big)^2/K^-_{\rm el}\ge 0$,  
otherwise $Q_{\rm el}=Q^{(1)}_{\rm el}$. The dependence of the functions in~\eqref{Static} 
on $E, \eta$ and $\eta^\prime$ is easily restored by looking at~\eqref{Splitting}.  

To solve the integral equation for $Q_{\rm ee}$, we employ an iterative approach. It expands
$Q_{\rm ee}$ in powers of $\tilde{K}^-_{\rm ee}$, that is, in the number of renormalized
backscattering events, as given in Eq.~\eqref{Qexpansion}. The expansion coefficients 
satisfy a set of linear integral equations. Writing out the dependencies on energies and 
direction cosines explicitly, the equations read 
\begin{widetext}
\begin{align}
Q_{\rm ee}^{(n)}(E|E^\prime;\eta\eta^\prime)&=\tilde{K}^{(n),-}_{\rm ee}(E|E^\prime;\eta\eta^\prime)
+ \!\int_{E^\prime}^E \!\!\!\! dE^{\prime\prime} \tilde{K}^+_{1,{\rm ee}}(E|E^{\prime\prime};E^\prime\eta\eta^\prime)
Q_{\rm ee}^{(n)}(E^{\prime\prime}|E^\prime;\eta\eta^\prime)
\nonumber\\
&+\int_{E^\prime}^E \!\! dE^{\prime\prime} Q_{\rm ee}^{(n)}(E|E^{\prime\prime};\eta\eta^\prime)
\tilde{K}^+_{2,{\rm ee}}(E^{\prime\prime}|E^\prime;E\eta\eta^\prime)~,~~n=1, 3, ...., n_\mathrm{max}
\label{Dynamic}
\end{align}
with
\begin{align}
\tilde{K}^{(2l+1),-}_{\rm ee}(E|E^\prime;\eta\eta^\prime)&= \sum_{k=0}^{l-1}
\int_{E^\prime}^E \!\!dE^{\prime\prime} \!\!\!\int_{E^\prime}^{E^{\prime\prime}}\!\!\!\!\!\! dE^{\prime\prime\prime}
Q_{\rm ee}^{(2k+1)}(E|E^{\prime\prime};\eta\eta^\prime)
\tilde{B}^-_{\rm ee}(E^{\prime\prime}|E^{\prime\prime\prime};EE^\prime\eta\eta^\prime)
Q_{\rm ee}^{(2(l-k)-1)}(E^{\prime\prime\prime}|E^\prime;\eta\eta^\prime)
\end{align}
\end{widetext}
for $l\ge 1$ and $\tilde{K}^{(1),-}_{\rm ee}=\tilde{K}^{-}_{\rm ee}$. By construction, the 
maximum number of backscattering events $n_\mathrm{max}$ is always odd. Due to 
the Volterra-type structure of the energy integrals, Eq.~\eqref{Dynamic} can be discretized in such a 
way that the $Q^{(k)}_{\rm el}$ appearing in the inhomogeneity $\tilde{K}^{(2l+1),-}_{\rm ee}$ are known from 
the previous steps of the calculation. We obtained convergence by iterating up to $n_\mathrm{max}=13$. 

In the numerical implementation we set an energy cut-off $E_\mathrm{max}=20\,\mathrm{eV}$ and discretized the 
energy interval $[-\Phi,E_\mathrm{max}]$ by $N=20$ slices. The interval of the internal direction cosines,
$[\eta_\mathrm{min},1]$, is also split into $M=20$ subintervals. Thus, after discretization, Eq.~\eqref{Dynamic}
is, for each iteration step, a linear algebraic equation on a $N\times N$ energy grid, which has to be solved
for each doublet $(\eta,\eta^\prime)$ drawn from the $M\times M$ angular grid. For the chosen discretization, 
the resolution in energy $\Delta E\approx 0.7\,\mathrm{eV}$, while the resolution in direction cosine 
$\Delta\eta \approx 0.05$. The integrals of the escape function~\eqref{EscapeFct} are calculated on the 
same grids, truncated, however, by the lower bounds $\eta_\mathrm{min}(E)$ and $E_\mathrm{min}(\eta^\prime)$, 
whereas the integrals required for $\gamma$, $K_1^+$, $K_2^+$, and $B^-$ (cf. Eqs.~\eqref{GammaFct}--\eqref{Bm}) 
are done by Gaussian integrations. For imperfect surfaces, finally, integrations over the external direction 
cosine $\xi$ are required, which again are discretized by $20$ slices. Numerically most expensive is the 
Monte-Carlo integration required for the electron-electron transition rate. The other parts of 
the code perform rather efficiently by taking advantage of the multicore structure of modern processors.

\section{Electron-electron transition rate}
\label{AppendixC}
The complete electron-electron transition rate, including direct and exchange scattering, 
on which the calculation of the emission yield is based, is given in this appendix. The 
manipulations presented below can be made for a dynamically screened Coulomb interaction 
$U(q,\omega)=1/q^2\varepsilon(q,\omega)$, but we give the result only for a statically 
screened Coulomb potential,
\begin{align}
U(q)=V(q,0)=\frac{1}{q^2+\kappa^2}~,
\end{align}
because this is the one we considered in the calculation of the emission yield
(cf. Eq.~\eqref{Vpseudo}). 

The starting point is the expression for the transition rate due to electron-electron
scattering~\cite{GL87}~,
\begin{align}
W_{\rm ee}(\vec{k}|\vec{k}^{\,\prime})&=
\frac{2}{\pi^3}\int d^3q\, d^3q^\prime n_{\rm F}(E_q)\bar{n}_{\rm F}(E_{q^\prime})
\nonumber\\
&\times |M(\vec{k}\vec{q}\,|\vec{k}^{\,\prime}\vec{q}^{\,\prime})|^2
\delta(E_k+E_q-E_{k^\prime}-E_{q^\prime})
\nonumber\\
&\times \delta(\vec{k}+\vec{q}-\vec{k}^{\,\prime}-\vec{q}^{\,\prime})~,
\label{WeeDXinitial}
\end{align}
with $\vec{k}$ and $\vec{k}^{\,\prime}$ the momenta of the electron in the initial and 
final state, $\bar{n}_{\rm F}(E_q)=1-n_{\rm F}(E_q)$, where $n_{\rm F}(E_q)$ is the Fermi 
function, and $M(\vec{k}\vec{q}\,|\vec{k}^{\,\prime}\vec{q}^{\,\prime})$ the sum of the 
three matrix elements corresponding to the three Coulomb terms diagrammatically shown 
in Fig.~\ref{MethodCartoon}c. Energies are measured from the bottom of the conduction band. 
Hence, for an effective electron mass equal to the bare electron mass, the situation to 
which we restrict our considerations, as mentioned in the main text, $E_k=k^2$ with 
$k=|\vec{k}|$ and similarly for the other energies. For the electrons of the Fermi sea, 
both spin orientations are taken into account.

The task is to express, after $\vec{q}{\,^\prime}$ is integrated out, 
the remaining momenta ($\vec{k}$,  $\vec{k}^{\,\prime}$ and $\vec{q}\,$) in terms of total 
energies ($E$, $E^\prime$, and $\tilde{E}$), direction cosines ($\eta$, $\eta^\prime$, and 
$\tilde{\eta}$), and azimuth angles ($\phi_k$, $\phi_{k\prime}$, and $\phi_q$). Distinguishing
forward and backward scattering with respect to the surface normal and integrating out the 
azimuth angles yields then the transition rate $W_{\rm ee}^\pm(E\eta|E^\prime\eta^\prime)$
needed for the kernels of the embedding equation and the calculation of $\gamma(E)$. In the 
following, the labels $p$, $p^\prime$, and $\tilde{p}$ denote, respectively, the sign of the 
z-components of the vectors $\vec{k}$, $\vec{k}^{\,\prime}$ and $\vec{q}$, while the variables 
$T$, $T^\prime$ and $\tilde{T}$ stand for the lateral energies associated with them. For instance, 
$T=(E+U)(1-\eta^2)$ and likewise for the other combinations. The lateral energies can thus be 
used as alternatives to the direction cosines. In terms of the functions which follow, 
\begin{align}
W^\pm_{\rm ee}(E\eta|E^\prime\eta^\prime)=W_{\rm ee}(ETp=1|E^\prime T^\prime p^\prime=\pm 1)~,
\end{align}
while the momentum transfer $g^\pm(E\eta|E^\prime\eta^\prime;\phi)$ employed in Eqs.~\eqref{WelInt} 
and~\eqref{WeeD} becomes
\begin{align}
g^\pm(E\eta|E^\prime\eta^\prime;\phi)=\tilde{g}(ETp=1|E^\prime T^\prime p^\prime=\pm 1;\phi)~.
\end{align}

Using the homogeneity in the lateral directions to measure the azimuth angles with respect to 
the projection of one of the momenta onto the $xy$-plane, for instance, the projection of the 
vector $\vec{k}$, and defining  
\begin{align}
R_1&=|\vec{k}-\vec{k}^{\,\prime}|_{pp^\prime}=\tilde{g}(ETp|E^\prime T^\prime p^\prime; \phi_{k^\prime})~,
\\
R_2&=|\vec{q}-\vec{k}^{\,\prime}|_{\tilde{p}p^\prime}=
\tilde{g}(\tilde{E}\tilde{T}\tilde{p}|E^\prime T^\prime p^\prime; \phi_q-\phi_{k^\prime})~,
\end{align}
where 
\begin{align}
\tilde{g}(&ETp|E^\prime T^\prime p^\prime;\phi)=\big(T+T^\prime - 2\sqrt{T T^\prime}\cos\phi 
\nonumber\\
&-[p\sqrt{E+U-T}-p^\prime\sqrt{E^\prime + U - T^\prime}]^2\big)^{1/2}~,
\end{align}
yields, after setting $\tilde{E}=E_q-U=\tilde{T}+\tilde{q}^2-U$ and working out the energy-conserving 
$\delta-$function, 
\begin{widetext}
\begin{align}
W_{\rm ee}(ETp|E^\prime T^\prime p^\prime)=\sum_{\tilde{p}=\pm 1}\, 
\int_0^\infty \!\!\!d\tilde{T}\int_0^{2\pi}\!\!\!d\phi_{k^\prime}\int_0^{2\pi}\!\!\! d\phi_{q}
\frac{U(R_1,R_2)N(E,E^\prime,\tilde{E})}{|r(ETp|E^\prime T^\prime p^\prime)|} 
\Theta\big(\tilde{q}(ETp|E^\prime T^\prime p^\prime;\tilde{p}\tilde{T}\phi_{k^\prime}\phi_{q})\big) 
\label{WeeDXfinal}
\end{align}
with $\Theta(x)=1$ for $x \geq 1$ and zero otherwise,  
\begin{align}
U(R_1,R_2)&=2\big( [U(R_1)]^2 + [U(R_2)]^2 - U(R_1)U(R_2) \big)~,
\label{U2fct}\\
N(E,E^\prime,\tilde{E})&= \pi^{-3} n_{\rm F}(\tilde{E}+U)\bar{n}_{\rm F}(E-E^\prime+\tilde{E}+ U)~,
\\
\tilde{q}(ETp|E^\prime T^\prime p^\prime;\tilde{p}\tilde{T}\phi_{k^\prime}\phi_{q})&=
\frac{s(ETp|E^\prime T^\prime p^\prime;\phi_{k^\prime})-h(T T^\prime;\phi_q,\phi_q-\phi_{k^\prime})\sqrt{\tilde{T}}}
{\tilde{p}\,r(ETp|E^\prime T^\prime p^\prime)}~,
\end{align}
where 
\begin{align}
s(ETp|E^\prime T^\prime p^\prime;\phi_{k^\prime}) &= 
E-E^\prime - [\tilde{g}(ETp|E^\prime T^\prime p^\prime;\phi_{k^\prime})]^2~, \\
h(T T^\prime; \phi_q,\phi_q-\phi_{k^\prime}) &= 2 (\sqrt{T}\cos\phi_q - \sqrt{T^\prime}\cos(\phi_q-\phi_{k^\prime})~,\\
r(ETp|E^\prime T^\prime p^\prime) &=2 (p\sqrt{E+U-T} - p^\prime \sqrt{E^\prime + U -T^\prime})~.
\end{align}
\end{widetext}

Although the expression for $W_{\rm ee}$ is perhaps somewhat involved, it follows straight from 
energy conservation, which after $\vec{q}^{\,\prime}$ is integrated out is encoded in 
$\delta(E_k+E_q-E_{k^\prime}-E_{|\vec{k}+\vec{q}-\vec{k}^{\,\prime}|})$. In terms of our variables, 
and suppressing all dependencies except the one on $\tilde{E}$, this becomes $\delta(f(\tilde{E}))$ 
with $f(\tilde{E})=s-h\sqrt{\tilde{T}}-\tilde{p} r \sqrt{\tilde{E}-\tilde{T}+U}$. Integrating 
over $\tilde{E}$ and renaming the solution $\tilde{E}_0$ of $f(\tilde{E}_0)=0$ by $\tilde{E}$ yields,
after taking the Jacobi determinant of the variable transformation into account, Eq.~\eqref{WeeDXfinal}.  

The remaining three integrals cannot be done analytically. We employed for it 
a Monte-Carlo integrator. Had we considered only the direct terms, a different strategy, based on rewriting 
the energy-conserving $\delta-$function in terms of an integral over a product of two $\delta-$ 
functions, could have been employed, leading to $W_{\rm ee}^\pm|_D$ given in Eq.~\eqref{WeeD}. In the 
notation of this appendix, the $W_{\rm ee}^\pm|_D$ arises from the $[U(R_1)]^2$ term in Eq.~\eqref{U2fct}. 
We verified that the result of the Monte-Carlo integration of this term alone coincides with the numerical 
result obtained directly from Eq.~\eqref{WeeD}. 

\bibliography{ref}
\bibliographystyle{apsrev}

\end{document}